\documentclass[prl,amsmath]{revtex4}  %% REVTeX 4.0
\usepackage{graphicx}% Include figure files
\usepackage{amssymb}
\usepackage{amsmath}

\begin{document}

%\begin{frontmatter}

\title{
On Optimal  Canonical Variables in the Theory of Ideal Fluid with
Free Surface}

\author{Pavel M. Lushnikov$^{1,2,3}   \quad and$
}
\author{Vladimir E. Zakharov$^{1,2}$
}

\affiliation{
  $^1$ Landau Institute for Theoretical Physics, Kosygin St. 2,
  Moscow, 119334, Russia
  \\
  $^2$ Department of Mathematics, University of Arizona, PO Box 210089, Tucson,
  Arizona, 85721
  \\
  $^3$ Department of Mathematics, University of  Notre Dame, Indiana 46556
  }

\email{zakharov@math.arizona.edu}

\date{October 24, 2004
}

\begin{abstract}
Dynamics of ideal fluid with free surface can be effectively
solved by perturbing the Hamiltonian in weak nonlinearity limit.
However it is shown that perturbation theory, which includes third
and fourth order terms in the Hamiltonian, results in the
ill-posed equations because of short wavelength instability. To
fix that problem we introduce the canonical Hamiltonian transform
from original physical variables to new variables for which
instability is absent.

\end{abstract}

%\begin{keyword}

%\pacs{}

\maketitle

\section{Introduction}

The Euler equations describing dynamics of ideal fluid with free
surface is a Hamiltonian system, which is especially simple if
the fluid motion is potential, ${\bf v}=\nabla \Phi$, where ${\bf
v}$ is the fluid's velocity and $\Psi$ is the velocity potential.
In this case \cite{Zakharov1966,Zakharov1968,ZakharovEurJ1999} the
Euler equations can be presented in the form:
\begin{equation}\label{hameq1}
\frac{\partial \eta}{\partial t}=\frac{\delta H}{\delta\Psi},
\quad \frac{\partial \Psi}{\partial t}=-\frac{\delta
H}{\delta\eta}.
\end{equation}
Here $z=\eta({\bf r})$ is the shape of surface, $z$ is vertical
coordinate and ${\bf r}=(x,y)$ are horizontal coordinates,
$\Psi\equiv \Phi\big |_{z=\eta}$ is the velocity potential on the
surface. The Hamiltonian $H$ coincides with the total (potential
and kinetic) energy of fluid. The Hamiltonian cannot be expressed
in a closed form as a function of surface variables $\eta,
\,\Psi$, but it can be presented by the infinite series in powers
of surface steepness $|\nabla \eta|$:
\begin{equation}\label{hamserdef}
H=H_0+H_1+H_2+\ldots.
\end{equation}
Here $H_0, \ H_1, \ H_2$ are quadratic, cubic and quartic terms,
respectively. Equations $(\ref{hameq1}),(\ref{hamserdef})$ are
widely used now for numerical simulation of the fluid dynamics
\cite{CraigWorfolk1995,DyachenkoKorotkevichZakharovJETPLett2003a,DyachenkoKorotkevichZakharovJETPLett2003b,DyachenkoKorotkevichZakharovPRL2004,OnoratoZakharovPRL2002,PushkarevEurJMech1996,PushkarevZakharovPRL1996,PushkarevZakharovPhysD2000,TanakaJFM2001,ZakharovDyachenkoVasilyevEurJ2002,Krasitskii1990}.
These simulations are performing by the use of the spectral code,
at the moment a typical grid is $512\times 512$ harmonics.

Canonical variables are used also for analytical study of the
surface dynamics in the limit of small steepness. It was shown
\cite{KuznetsovSpektorZakharov1993,KuznetsovSpektorZakharov1994,DyachenkoZakharovKuznetsov1996}
that the simplest truncation of the series $(\ref{hamserdef})$,
namely
\begin{equation}\label{htrunc1}
H=H_0+H_1,
\end{equation}
leads to completely integrable model - complex Hopf equation. In
framework of this approach one can develop the self-consistent
theory of singularity formation in absence of gravity and
capillarity for two dimensions (one vertical coordinate $z$ and
one horizontal coordinate $x$).

However, use of canonical variables $\eta, \, \Psi$ has a weak
point, which becomes clear, if we concentrate our attention on the
complex Hopf equation,
\begin{equation}\label{hopfp}
\frac{\partial \Pi}{\partial t}=-\frac{1}{2}\Big (\frac{\partial
\Pi}{\partial x} \Big )^2,
\end{equation}
which comes from Eqs. $(\ref{hameq1}),(\ref{htrunc1})$. Here
\begin{equation}\label{psipm}
\Psi=Re(\Pi)
\end{equation}
and $\Pi$ is the analytic function of the complex variable $x$ in
a strip $ -h \leq Im(x)\leq 0$, $h$ is the depth of the fluid. The
weak point is that Eq. $(\ref{hopfp})$ is ill-posed. A general
complex solution of this equation is unstable with respect to
grow of small short-wave perturbations. The same statement is
correct with respect to more exact fourth order Hamiltonian
\begin{equation}\label{hamfourth}
H=H_0+H_1+H_2,
\end{equation}
which is used in most numerical experiments.  These experiments
are  easy becomes unstable: to arrest instability one should
include into equations strong artificial damping at high wave
numbers. Even in presence of such damping one can simulate only
waves of a relatively small steepness (not more than
$0.15$).%[reference??????????].

In this Article we show that these difficulties can be fixed by a
proper canonical transformation to another canonical variables.
It is remarkable, but the property of nonlinear wave equation to
be well- or ill-posed is {\it not invariant with respect to
choice of the variables}.

In the present Article we demonstrate that there are new canonical
variables such that the Eqs. $(\ref{hameq1}),(\ref{hamfourth})$
are well-posed if we consider the nonlinearity up to the fourth
order in the Hamiltonian. We call these variables ``optimal
canonical variables". We demonstrate in the present Article that
the choice of the optimal canonical variables is unique  provided
we additionally require the Hamiltonian system to be free of short
wavelength instability for largest possible steepness of the
surface, i.e. for the largest possible nonlinearity. We
conjecture  that optimal canonical variables allow simulation
with higher steepness compare  with standard variables $\Psi, \,
\eta.$ We can also formulate a conjecture that the optimal
canonical variables exist  in all orders of nonlinearity.

\section{Basic equations and the Hamiltonian formalizm}

Consider the dynamics of an incompressible ideal fluid with free
surface and constant depth. Fluid occupies the region
\begin{equation}\label{etadef}
-h<z <\eta({\bf r}), \quad {\bf r}=(x,y),
\end{equation}
where $(x,y)$ are the horizontal coordinates and $z$ is the
vertical coordinate.

Viscosity is assumed to be absent and the fluid's velocity ${\bf
v}$ is potential one,
\begin{equation}\label{phidef}
{\bf v}=\nabla \Phi,
\end{equation}
where $\Phi$ is the velocity potential. Incompressibility
condition,
\begin{equation}\label{incompr}
\nabla \cdot{\bf v}=0,
\end{equation}
results in the Laplace Eq.
\begin{equation}\label{lapl1}
\triangle\Phi=0.
\end{equation}

The potential $\Phi$ satisfies also the Bernoulli equation:
\begin{equation}\label{bernoulli}
\Phi_t+\frac{1}{2}\big (\nabla \Phi\big )^2+p+gz=0,
\end{equation}
where $p$ is the pressure, $g$ is the acceleration of gravity,
and we set density of fluid to unity.

There are two types of boundary conditions at free surface for
Eqs. $(\ref{lapl1}),(\ref{bernoulli})$. First is the  kinematic
boundary condition
\begin{equation}\label{kinem1}
\frac{\partial \eta}{\partial t}=\Big (\Phi_z-\nabla\eta\cdot
\nabla \Phi\Big )\Big|_{z=\eta}=v_n\sqrt{1+(\nabla \eta)^2},
\end{equation}
where $v_n={\bf n}\cdot \nabla \Phi$ is the normal component of
fluid's velocity at free surface, and ${\bf n}=(-\nabla\eta,1)\big
[1+(\nabla\eta)^2\big ]^{-1/2}$
 is the interface normal vector.

Second is the dynamic boundary condition at free surface
\begin{equation}\label{dynam1}
p \big |_{z=\eta}=\sigma \nabla \cdot \frac{\nabla
\eta}{\sqrt{1+(\nabla\eta)^2}},
\end{equation}
where $\sigma$ is the surface tension coefficient which
determines the jump of the pressure at free surface from zero
value outside of the fluid to $p \big |_{z=\eta}$ value inside
fluid.

Boundary condition at the bottom is
\begin{equation}\label{boundarybottom1}
 \Phi_z|_{z=-h}=0.
\end{equation}

Eqs. $(\ref{lapl1})-(\ref{boundarybottom1})$ form a closed set of
equations to determine the dynamics of free surface.

 The total energy, $H$, of the fluid consists of the kinetic energy,
$T,$ and the potential energy, $U$:
\begin{eqnarray}\label{hamdef}
H=T+U, \\
\label{tdef}T=\frac{1}{2}\int d{\bf r}\int^\eta_{-h} \big (\nabla
\Phi\big)^2
dz, \\
\label{udef}U=\frac{1}{2}g\int \eta^2\, d{\bf r}+\sigma\int \Big
[\sqrt{1+\big(\nabla \eta\big )^2}-1\Big ]d{\bf r}.
\end{eqnarray}

It is convenient to introduce the value of the velocity potential
at interface as
\begin{equation}\label{psidef0}
\Phi\big |_{z=\eta}\equiv \Psi({\bf r},t).
\end{equation}
It was shown in Ref. \cite{Zakharov1968} that the free surface
problem $(\ref{lapl1})-(\ref{boundarybottom1})$ can be written in
the Hamiltonian form $(\ref{hameq1})$, with the Hamiltonian $H$
defined in $(\ref{hamdef})$.

Fourier transform,
\begin{equation}\label{four1}
\Psi_{{\bf k}}=\frac{1}{2\pi}\int\exp(-i{\bf k}\cdot {\bf r})\Psi
({\bf r}) d {\bf r},
\end{equation}
is the canonical transform which conserves the Hamiltonian
structure and Eqs. $(\ref{hameq1})$ take the following form:
\begin{equation}\label{hamfoureq1}
\frac{\partial \eta_{\bf k}}{\partial t}=\frac{\delta
H}{\delta\Psi_{-{\bf k}}}, \quad \frac{\partial \Psi_{\bf
k}}{\partial t}=-\frac{\delta H}{\delta\eta_{-{\bf k}}},\quad
\eta_{\bf k}^*=\eta_{-{\bf k}}, \quad \Psi_{\bf k}^*=\Psi_{-{\bf
k}}.
\end{equation}

\section{Weak nonlinearity}

If a typical slope of free surface is small, $|\nabla \eta|\ll 1$,
the Hamiltonian $H$ can be series expanded (see Eq.
$(\ref{hamserdef})$) in powers of steepness $|\nabla \eta|$ which
gives \cite{Zakharov1968,ZakharovEurJ1999}:
\begin{eqnarray}%\label{hamser}
%H=H_0+H_1+H_2+\ldots, \\
\label{ham0} H_0=\frac{1}{2}\int \Big \{A_k |\Psi_{\bf k}|^2+B_k
|\eta_{\bf k}|^2 \Big
\}d{\bf k}, \nonumber \\
 A_k=k\tanh(kh), \ B_k=g+\sigma k^2, \ k=|{\bf k}|,\\
\label{ham1} H_1=\frac{1}{4\pi}\int L^{(1)}_{{\bf k}_1,{\bf
k}_2}\Psi_{{\bf k}_1}\Psi_{{\bf k}_2}\eta_{{\bf k}_3}\delta({\bf
k}_1+{\bf
k}_2+{\bf k}_3)d{\bf k}_1d{\bf k}_2d{\bf k}_3 \\
\label{ham2} H_2=\frac{1}{2(2\pi)^2}\int \Big [L^{(2)}_{{\bf
k}_1,{\bf k}_2,{\bf k}_3,{\bf k}_4}\Psi_{{\bf k}_1}\Psi_{{\bf
k}_2}-\frac{\sigma}{4}  ({\bf k}_1\cdot{\bf k}_2)({\bf
k}_3\cdot{\bf k}_4)\eta_{{\bf k}_1}\eta_{{\bf k}_2}\Big ]
 \nonumber \\
\times \eta_{{\bf k}_3}\eta_{{\bf k}_4}\delta({\bf k}_1+{\bf
k}_2+{\bf k}_3+{\bf k}_4)d{\bf k}_1d{\bf k}_2d{\bf k}_3d{\bf k}_4,
\end{eqnarray}
where matrix elements are given by
\begin{eqnarray}\label{L1L2def}
L^{(1)}_{{\bf k}_1,{\bf k}_2}=-{\bf k}_1\cdot{\bf
k}_2-A_{1}A_{2}, \nonumber \\
L^{(2)}_{{\bf k}_1,{\bf k}_2,{\bf k}_3,{\bf k}_4}=\frac{1}{4}A_1
A_2\Big(A_{1+3}+A_{2+3}+A_{1+4}+A_{2+4}\Big )\nonumber
\\
-\frac{1}{2}(k_1^2A_2+k_2^2A_1), \quad A_j\equiv A_{k_j}, \
A_{j+l}\equiv A_{{\bf k}_j+{\bf k}_l}.
\end{eqnarray}

The corresponding dynamical equations follow from
$(\ref{hameq1}),(\ref{hamfourth}),(\ref{ham0}),(\ref{ham1}),(\ref{ham2}):$
\begin{eqnarray}\label{dynamics1}
\frac{\partial \Psi}{\partial t}=-g\eta+\sigma \triangle
\eta+\frac{1}{2}\Big[\big (\hat A\Psi\big )^2-\big (\nabla
\Psi\big )^2\Big ]-(\hat A\Psi)\hat A\big[\eta(\hat A\Psi)\big ]\nonumber\\
-(\triangle \Psi)(\hat A\Psi)\eta-\frac{\sigma}{2}\nabla\cdot\Big
[\nabla \eta(\nabla \eta\cdot\nabla \eta)\Big ], \nonumber \\
\frac{\partial \eta}{\partial t}=\hat A \Psi-\nabla \cdot \big [
(\nabla \Psi)\eta\big ]-\hat A \big[\eta\hat A\Psi\big ]+\hat
A\Big \{\eta\hat A\big [\eta\hat A \Psi\big ]\Big \}\nonumber\\
+\frac{1}{2}\triangle \big [\eta^2\hat A\Psi\big
]+\frac{1}{2}\hat A\big [\eta^2\triangle \Psi\big ],
\end{eqnarray}
where $\hat A$ is the linear integral operator which corresponds
to multiplication on $k\tanh(kh)$ in Fourier space. For two
dimensional flow, $\Psi(x,y)=\Psi(x), \, \eta(x,y)=\eta(x)$, this
operator is given by
\begin{eqnarray}\label{Adef}
\hat A=-\frac{\partial }{\partial x}\hat
R\\
\label{Rdef} \hat Rf(x)=\frac{1}{2 h} P.V.
\int^{+\infty}_{-\infty}\frac{f(x')}{\sinh\big [
(x'-x)\pi/(2h)\big ]} \ dx'
\end{eqnarray}
 where $P.V.$ means Cauchy principal value of
integral. In the limiting case of infinitely deep water, $h\to
\infty$, the operator   $\hat A$ turns into the operator $\hat k$
\begin{equation}\label{kdef}
\lim_{h\to \infty}\hat A=\hat k
\end{equation}

 which
corresponds to multiplication on $|{\bf k}|$ in Fourier space
while  the operator $\hat R$ for two-dimentsional flow turns into
the Hilbert transform:
\begin{equation}\label{hilb}
\lim_{h\to \infty}\hat R=\hat H, \quad \hat Hf(x)=\frac{1}{\pi}
P.V. \int^{+\infty}_{-\infty}\frac{f(x')}{x'-x}dx'.
\end{equation}
$\hat H$ can be also interpreted as a Fourier transform of $i \,
\mbox{sign}(k)$.

If one neglects gravity and surface tension, $g=0, \, \sigma=0,$
then Eqs. $(\ref{hameq1}),(\ref{hamserdef})$, at leading order
over small parameter $|\nabla \eta|$, result
in\cite{KuznetsovSpektorZakharov1994,KuznetsovSpektorZakharov1993,DyachenkoZakharovKuznetsov1996}
\begin{subequations}
\begin{eqnarray}\label{etaeq0}
\frac{\partial\eta}{\partial t}=\hat A \Psi,
\\
\label{psieq0} \frac{\partial \Psi}{\partial
t}=\frac{1}{2}\Big[\big (\hat A\Psi\big )^2-\big (\nabla \Psi\big
)^2\Big ].
\end{eqnarray}
\end{subequations}

Remarkable feature of Eqs. $(\ref{etaeq0}),(\ref{psieq0})$ is
that the second Eq. $(\ref{psieq0})$ does not depend on $\eta$
thus one can first solve  $(\ref{psieq0})$ and then find $\eta$
from Eq. $(\ref{etaeq0})$. Substitution $\Pi \equiv\Psi+i\hat
R\Psi$ into Eq. $(\ref{psieq0})$ results in the complex Hopf Eq.
$(\ref{hopfp})$ for two-dimensional flow
\cite{DyachenkoZakharovKuznetsov1996} which is completely
integrable.

Both Eqs. $(\ref{psieq0})$ and $(\ref{hopfp})$  are ill-posed
because they have short wavelength instability which is determined
as follows: we can analyze Eq. $(\ref{psieq0})$ and take $\Psi$ in
the form
\begin{eqnarray}\label{psilin1}
\Psi=\Psi_0+\Big (\Psi_1 e^{i{\bf k}\cdot {\bf r}+\nu t}+c.c.\Big
),
\end{eqnarray}
where $\Psi_0({\bf r},t)$ is a solution of Eq. $(\ref{psieq0})$,
$\Psi_1$ is the amplitude of small perturbation, and {\it c.c.}
means complex conjugation. Then, in the limit $|{\bf k}|\to
\infty$, $\Psi_0$ evolves very slow in space compare to $e^{i{\bf
k}\cdot {\bf r}+\nu t}$ and we get the dispersion relation for
small perturbations:
\begin{equation}\label{nu01}
\nu=A_{k}\hat A\Psi_0-i{\bf k}\cdot\nabla\Psi_0
\end{equation}
which describes instability for $Re(\nu)=A_{k}\hat A\Psi_0>0$.
For general initial condition such instability region always
exists. The instability growth rate, $Re(\nu)$ grows as $|{\bf
k}|$ increases.

\section{Short wavelength stability analysis of the fourth-order Hamiltonian}

To study linear stability of the Hamiltonian system in respect to
short wavelength perturbations one can set
\begin{eqnarray}\label{etapsienvelope1}
\eta_{\bf k}=\eta_{0 \, {\bf k}}+\delta \eta_{\bf k}, \nonumber\\
\Psi_{\bf k}=\Psi_{0 \, {\bf k}}+\delta \Psi_{\bf k},
\end{eqnarray}
where $\eta_{0 \, {\bf k}},\, \Psi_{0 \, {\bf k}}$ are solutions
of Eqs. $(\ref{hameq1}),(\ref{hamfourth})$ and $\delta \eta_{
{\bf k}},\, \delta \psi_{{\bf k}}$ are short wavelength
perturbations localized around wave vector ${\bf k}, |{\bf k}
|\gg q$, $q$ is a typical wavenumber for $\eta_{0 \, {\bf k}},\,
\Psi_{0 \, {\bf k}}$.

If we take into account contribution to the fourth-order
Hamiltonan up to second order in amplitude of perturbations
$\delta \eta_{ {\bf k}},\, \delta \Psi_{{\bf k}}$ we get the
following general form of the perturbed Hamiltonian:
\begin{eqnarray}\label{H2lin}
\delta H_0=\frac{1}{2} \int \tilde A_{\bf k}|\delta\Psi_{\bf
k}|^2d{\bf k}+\frac{1}{2} \int \tilde B_{\bf k}|\delta\eta_{\bf
k}|^2 d{\bf k}+ \int (F_{\bf k}+iG_{\bf k})\delta\Psi_{\bf k}
\delta\eta_{-{\bf k}} d{\bf k},
\nonumber\\
\tilde A_{\bf k}=\tilde A_{-{\bf k}},\ \tilde B_{\bf k}=\tilde
B_{-{\bf k}}, \ F_{\bf k}=F_{-{\bf k}}, \ G_{\bf k}=-G_{-{\bf k}},
\end{eqnarray}
where $\tilde A_{\bf k}, \ \tilde B_{-{\bf k}}, \ F_{\bf k}, \
G_{\bf k}$ are real and depend on $\eta_{0 \, {\bf k}},\, \Psi_{0
\, {\bf k}}$. Here we disregard linear contribution to $\delta
H_0$ because it has no effect on linear stability analysis.

It follows from Eqs. $(\ref{hamfoureq1}),(\ref{H2lin})$ that
equations of motion take the following form:
\begin{eqnarray}\label{H2linEq}
\frac{\partial\delta\eta_{\bf k}}{\partial t}=
 \tilde A_{\bf k}\delta\Psi_{\bf
k}+(F_{\bf k}-iG_{\bf k})\eta_{\bf k}, \nonumber \\
\frac{\partial\delta\Psi_{\bf k}}{\partial t}=
 -\tilde B_{\bf k}\delta\eta_{\bf
k}-(F_{\bf k}+iG_{\bf k})\Psi_{\bf k}.
\end{eqnarray}
An assumption of exponential dependence on time,
\begin{eqnarray}\label{psilink2}
\delta\eta_{\bf k} \sim  e^{\nu_{\bf k} t}, \ \delta\Psi_{\bf
k}\sim  e^{\nu_{\bf k} t},
\end{eqnarray}
gives a dispersion relation
\begin{eqnarray}\label{nuk2}
\nu_{\bf k} =-iG_{\bf k}\pm \sqrt{F_{\bf k}^2-\tilde A_{\bf
k}\tilde B_{\bf k}}
\end{eqnarray}
which describes instability provided $F_{\bf k}^2-\tilde A_{\bf
k}\tilde B_{\bf k}>0.$

\section{Ill-posedness of the fourth-order Hamiltonian}

Consider now a general case of nonzero $g$ and $\sigma$ and take
into account all terms in the Hamiltonian up to forth order, i.e.
consider full Eqs. $(\ref{dynamics1})$. At the leading order over
steepness $\Theta$ and wavenumber $k$ we obtain:
\begin{eqnarray}\label{AB2}
\tilde A_{\bf
k}=A_k+(k^2-A_k^2)\eta_0-A_k(k^2-A_k^2)\eta_0^2+O(k\Theta^2),
\nonumber\\
\tilde B_{\bf k}=B_k+ A_k(\hat A\Psi_0)^2+O( k^{0}v_0^2l_0^{-1}), \nonumber \\
F_{\bf k}=-A_k\hat A\Psi_0- (k^2-A_k^2)(\hat
A\Psi_0)\eta_0+A_k\big (\hat A\big [\eta_0\hat A\Psi_0\big
]+\eta_0\nabla^2 \Psi_0\big ) \nonumber \\
+ O(k^0v_0 l_0^{-1}),\nonumber\\
G_{\bf k}={\bf k}\cdot\nabla \Psi_0+O(kv_0\Theta),
\end{eqnarray}
where $\eta_0=\frac{1}{2\pi}\int \eta_{\bf k} d{\bf k}, \
\Psi_0=\frac{1}{2\pi}\int \Psi_{\bf k} d{\bf k},$ and the
steepness is defined as $\Theta\sim |\nabla \eta_0|$. We
introduced here the typical value of fluid velocity,
$v_0\sim|\nabla \Psi_0|$ and the typical scale, $l_0$, of
variation of $v_0$ and $\eta_0:$ $\Theta\sim \eta_0/l_0, \
|\nabla v_0|\sim v_0/l_0.$

Eqs. $(\ref{nuk2}),(\ref{AB2})$ give instability growth rate. We
consider particular cases. If $\sigma \neq 0$ then in the limit
$|{\bf k}|\to \infty$, we have
\begin{eqnarray}\label{nuksigma2}
\nu_{\bf k} =\pm i\sqrt{\sigma k^3},
\end{eqnarray}
i.e. instability is absent. In derivation of Eq.
$(\ref{nuksigma2})$ we used exponential smallness of expression
$k^2-A_k^2=k^2/(\cosh{kh})^2\simeq 4k^2\exp(-kh)\ll 4k^2$ because
limit $|{\bf k}|\to \infty$ implies $k h\gg 1$. Thus finite
$\sigma$ makes problem
$(\ref{hamfoureq1}),(\ref{ham0})-(\ref{ham2})$ well-posed.

Note that for finite depth $A_k<k$ we could still have
instability at finite range of wavenumbers $kh\sim 1$. In that
case $k\eta_0\sim \eta_0/h\ll 1$ because a typical variation of
surface elevation, $\eta_0$, should be small to allow  weak
nonlinearity approximation used throughout this Article. Because
$k\eta_0\ll 1$, Eqs. $(\ref{nuk2}),(\ref{AB2})$ are reduced to
\begin{eqnarray}\label{nuksigmah2}
\nu_{\bf k}=-i{\bf k}\cdot\nabla \Psi_0\pm A_k^{1/2} \big
[-B_k+2(k^2-A_k^2)(\hat A\Psi_0)^2\eta_0\big ]^{1/2},  \quad
kh\sim 1,
\end{eqnarray}
which gives instability provided
\begin{eqnarray}\label{instabcond01}
B_k<2(k^2-A_k^2)(\hat A\Psi_0)^2\eta_0, \quad kh\sim 1.
\end{eqnarray}
E.g. instability occurs for $g=\sigma=0$:
\begin{eqnarray}\label{nukh2}
\nu_{\bf k} =-i{\bf k}\cdot\nabla \Psi_0\pm k (\hat
A\Psi_0)\eta_0 \sqrt{k^2-A_k^2}, \quad kh\sim 1.
\end{eqnarray}

We can estimate inequality $(\ref{instabcond01})$ as
\begin{eqnarray}\label{instabestimate01}
(g+\frac{\sigma}{h^2})\lesssim \frac{v_0^2\eta_0}{h^2},
\end{eqnarray}
where $v_0\sim|\nabla \Psi_0|$ is the typical velocity of fluid.

It follows from $(\ref{instabestimate01})$ that instability
occurs for large values of $v_0$ (because $\eta_0$ is small). If
gravity dominates, $g>\sigma/h^2$, then $(\ref{instabestimate01})$
gives $gh/v_0^2\lesssim \eta_0/h$
 but weak nonlinearity approximation implies that $\eta_0/h\ll 1$ which indicates that the kinetic energy
 strongly exceed the potential energy. Fluid has enough  kinetic energy to easily move upward at distance $\sim h$.
 As a result, at
 later stage of evolution weak nonlinearity approximation is
 violated and surface is strongly perturbed at scales $\sim h$.

 If capillarity dominates, $g<\sigma/h^2,$ inequality $(\ref{instabestimate01})$ gives $\sigma/(v_0^2\eta_0)\lesssim
 1$ and the kinetic energy again strongly exceed the potential energy.
 Assume now that, because of instability for $kh\sim 1$,  at later time of evolution the potential
 energy will be of the  same order as the kinetic energy, namely,
 $\eta_0 v_0^2\sim \sigma \Theta^2, \ \Theta\sim |\nabla \eta|$.
 Then $\sigma/(v_0^2\eta_0)\lesssim
 1$ results in inequality $\Theta \gtrsim 1$ which again violates weak
 nonlinearity approximation. Thus for arbitrary relations between $g$ and $\sigma/h^2$,
 and for $kh\sim 1$, the instability is
 possible for strong enough velocity of fluid and this instability
 results in violation of weak nonlinearity approximation in
 course of fluid evolution. In that sense there
 is no surprise that for large velocity there is an instability for $kh \sim 1$.
 This instability is purely physical which leaves problem well-posed.

Outside capillary scale we can set $\sigma=0$ and get from Eqs.
$(\ref{nuk2}),(\ref{AB2})$ that zero capillarity makes Eqs.
$(\ref{dynamics1})$ ill-posed for $k\to \infty:$
\begin{eqnarray}\label{nukhzerosigma2}
\nu_{\bf k} =-i{\bf k}\cdot\nabla \Psi_0\pm 2^{1/2}(\hat
A\Psi_0)^{1/2}k \big (\hat A\big [\eta_0\hat A\Psi_0\big
]+\eta_0\nabla^2 \Psi_0\big )^{1/2} \nonumber \\
 \sim -i{\bf
k}\cdot\nabla \Psi_0\pm kv_0\Theta^{1/2}.
\end{eqnarray}
An expression $\hat A\Psi_0\Big (\hat A\big [\eta_0\hat
A\Psi_0\big ]+\eta_0\nabla^2 \Psi_0\Big )$ in Eq.
$(\ref{nukhzerosigma2})$ is not sign-definite which results in
instability of the system
$(\ref{hamfoureq1}),(\ref{ham0})-(\ref{ham2})$. It means that the
fourth order Hamiltonian does not prevent short-wavelength
instability but makes instability weaker by the small factor
$\Theta^{1/2}$ compare with instability of the third-order
Hamiltonian (compare Eqs. $(\ref{nukhzerosigma2})$ and
$(\ref{nu01})$). Instability $(\ref{nukhzerosigma2})$ has been
observed numerically \cite{Dyachenkounpublished}. We conclude
that full fourth order system
$(\ref{hamfoureq1}),(\ref{ham0})-(\ref{ham2})$ is
 ill-posed for zero capillarity, $\sigma=0.$

Ill-posedness of Eqs.
$(\ref{hamfoureq1}),(\ref{ham0})-(\ref{ham2})$ can be also
interpreted as violation of perturbation expansion
$(\ref{hamserdef})$ for $k\to\infty.$ Namely,  short wavelength
contribution to the quadratic Hamiltonian
$(\ref{H2lin}),(\ref{AB2})$ is not small compare with the other
terms in the Hamiltonian $(\ref{ham0})-(\ref{ham2})$ provided
$k\eta_0\gtrsim 1.$

Ill-posedness makes Eqs.
$(\ref{hamfoureq1}),(\ref{ham0})-(\ref{ham2})$ (or, equivalently,
Eqs. $(\ref{dynamics1})$) difficult for simulations. There a few
ways to cope with that problem. One way is to resolve all scales
down to capillary scales which is extremely costly numerically.
E.g., if we want to study water waves in gravitation region
(scale of meters and larger), we would have to simultaneously
resolve capillary scale $\sim 1cm$. Other way is to introduce
artificial damping for short wavelengths, i.e. to replace Eqs.
$(\ref{hamfoureq1})$ by
\begin{equation}\label{hamfoureqregularazied1}
\frac{\partial \eta_{\bf k}}{\partial t}=\frac{\delta
H}{\delta\Psi_{-{\bf k}}}+\gamma_1(k)\eta_{\bf k}, \quad
\frac{\partial \Psi_{\bf k}}{\partial t}=-\frac{\delta
H}{\delta\eta_{-{\bf k}}}+\gamma_2(k)\Psi_{\bf k},
\end{equation}
where functions $\gamma_1(k), \,\gamma_2(k)$ are zero for small
and intermediate values of $k$ but they tend to $-\infty$ for
$k\to \infty$. Also it is possible to introduce finite viscosity
of the fluid. However in that case we would have to resolve very
small scales and, in addition, the Hamiltonian is not conserved
for finite viscosity so that we can not use the Hamiltonian
formalism.

In this Paper we use another way which is to  completely remove
short wavelength instabilities and make problem well-posed by
appropriate canonical transform from variables $\eta, \, \Psi$ to
new canonical variables $\xi, \, R.$

\section{Canonical transform}

Canonical transform from variables $\Psi, \eta$ to new variables
$R,\xi$ is determined by a generating functional $S:$
\begin{eqnarray}\label{Sdef}
S=\int R_{\bf k}\eta_{-{\bf k}}d{\bf k}+\frac{1}{8\pi}\int
A_3\eta_{{\bf k}_1}\eta_{{\bf k}_2}R_{{\bf k}_3}\delta({\bf
k}_1+{\bf k}_2+{\bf k}_3)d{\bf
k}_1d{\bf k}_2d{\bf k}_3   \nonumber \\
+\frac{1}{4(2\pi)^2}\int V_{{\bf k}_1,{\bf k}_2,{\bf k}_3,{\bf
k}_4}R_{{\bf k}_1}\eta_{{\bf k}_2}\eta_{{\bf k}_3}\eta_{{\bf
k}_4}\delta({\bf k}_1+{\bf k}_2+{\bf k}_3+{\bf k}_4)d{\bf
k}_1d{\bf k}_2d{\bf k}_3d{\bf k}_4,
\end{eqnarray}
\begin{subequations}
\begin{eqnarray}\label{psidef}
 \Psi_{\bf k}=\frac{\delta S}{\delta \eta_{-{\bf
k}}}=R_{\bf k}+\frac{1}{4\pi} \int A_1 R_{{\bf k}_1}\eta_{{\bf
k}_2}\delta({\bf k}_1+{\bf k}_2-{\bf k})d{\bf k}_1d{\bf k}_2
\nonumber \\
+\frac{3}{4(2\pi)^2}\int V_{{\bf k}_1,{\bf k}_2,{\bf k}_3,-{\bf
k}}R_{{\bf k}_1}\eta_{{\bf k}_2}\eta_{{\bf k}_3}\delta({\bf
k}_1+{\bf k}_2+{\bf k}_3-{\bf k})d{\bf k}_1d{\bf k}_2d{\bf k}_3,
\\
\label{xidef}\xi_{\bf k}=\frac{\delta S}{\delta R_{-{\bf
k}}}=\eta_{\bf k}+\frac{1}{8\pi} \int A_k \eta_{{\bf
k}_1}\eta_{{\bf k}_2}\delta({\bf k}_1+{\bf k}_2-{\bf k})d{\bf
k}_1d{\bf k}_2
 \nonumber \\
+\frac{1}{4(2\pi)^2}\int V_{-{\bf k},{\bf k}_2,{\bf k}_3,{\bf
k}_4}\eta_{{\bf k}_2}\eta_{{\bf k}_3}\eta_{{\bf k}_4}\delta({\bf
k}_2+{\bf k}_3+{\bf k}_4-{\bf k})d{\bf k}_2d{\bf k}_3d{\bf k}_4,
\end{eqnarray}
\end{subequations}
where $V_{{\bf k}_1,{\bf k}_2,{\bf k}_3,{\bf k}_4}$ is the
symmetric function of ${\bf k}_2,{\bf k}_3,{\bf k}_4.$ This is
the most general form of canonical transform up to terms of the
fourth order. The only condition which we use here is that $S$ is
chosen to be linear functional of $R$ to preserve the quadratic
dependence of the Hamiltonian on canonical momentum $R$.

The quantity $\eta$ can be found from Eq. $(\ref{xidef})$ as the
functional of $\xi$ by iterations (here and below we take into
account only corrections up to the fourth order in the
Hamiltonian):
\begin{eqnarray}\label{etaeq1}
\eta_{\bf k}=\xi_{\bf k}-\frac{1}{8\pi} \int A_k \xi_{{\bf
k}_1}\xi_{{\bf k}_2}\delta({\bf k}_1+{\bf k}_2-{\bf k})d{\bf
k}_1d{\bf k}_2
+\frac{1}{8(2\pi)^2} \nonumber \\
\times \int \Big [A_k A_{1+2}-2V_{-{\bf k},{\bf k}_1,{\bf
k}_2,{\bf k}_3}\Big ]\xi_{{\bf k}_1}\xi_{{\bf k}_2}\xi_{{\bf
k}_3}\delta({\bf k}_1+{\bf k}_2+{\bf k}_3-{\bf k})d{\bf k}_1d{\bf
k}_2d{\bf k}_3,
\end{eqnarray}
Eqs. $(\ref{psidef}),(\ref{etaeq1})$ give:
\begin{eqnarray}\label{psieq1}
 \Psi_{\bf
k}=R_{\bf k}+\frac{1}{4\pi} \int A_1 R_{{\bf k}_1}\xi_{{\bf
k}_2}\delta({\bf k}_1+{\bf k}_2-{\bf k})d{\bf k}_1d{\bf k}_2 +
\nonumber \\ \frac{1}{8(2\pi)^2} \int \Big [-A_1A_{2+3} +6V_{{\bf
k}_1,{\bf k}_2,{\bf k}_3,-{\bf k}}\Big ] R_{{\bf k}_1}\xi_{{\bf
k}_2}\xi_{{\bf k}_3}\nonumber
\\ \times \delta({\bf k}_1+{\bf k}_2+{\bf k}_3-{\bf
k})d{\bf k}_1d{\bf k}_2d{\bf k}_3.
\end{eqnarray}

Using  Eqs.
$(\ref{ham0}),(\ref{ham1}),(\ref{ham2}),(\ref{psidef}),(\ref{etaeq1})$
we get:
\begin{eqnarray}
\label{hamxi0} H_0=\frac{1}{2}\int \Big \{A_k |R_{\bf k}|^2+B_k
|\xi_{\bf k}|^2 \Big
\}d{\bf k}, \\
\label{hamxi1} H_1=\frac{1}{4\pi}\int \Big[-({\bf k}_1\cdot{\bf
k}_2)R_{{\bf k}_1}R_{{\bf
k}_2}-\frac{1}{6}(A_1B_1+A_2B_2+A_3B_3)\xi_{{\bf k}_1}\xi_{{\bf
k}_2}\Big ] \nonumber\\
\times\xi_{{\bf k}_3}\delta({\bf k}_1+{\bf
k}_2+{\bf k}_3)d{\bf k}_1d{\bf k}_2d{\bf k}_3, \\
H_2=\frac{1}{8(2\pi)^2}\int \Big \{ ({\bf k}_1\cdot{\bf
k}_2)(A_{1+2}-A_1-A_2)-k_1^2A_2-k_2^2A_1
\nonumber\\+\frac{1}{4}A_1A_2[A_{1+3}+A_{2+3}+A_{1+4}+A_{2+4}]
 +3[A_1V_{{\bf k}_2,{\bf k}_3,{\bf k}_4,{\bf k}_1}
 \nonumber \\
+A_2V_{{\bf k}_1,{\bf k}_3,{\bf k}_4,{\bf k}_2}] \Big \}
%\nonumber\\ \times
 R_{{\bf k}_1}R_{{\bf k}_2}\xi_{{\bf
k}_3}\xi_{{\bf k}_4}\delta({\bf k}_1+{\bf k}_2+{\bf k}_3+{\bf
k}_4)d{\bf k}_1d{\bf k}_2d{\bf k}_3d{\bf k}_4
\nonumber \\
\label{hamxi2} +\frac{1}{8(2\pi)^2}\int \Big \{ -\sigma({\bf
k}_1\cdot{\bf k}_2)({\bf k}_3\cdot{\bf
k}_4)+\frac{1}{4}A_{1+2}^2B_{1+2}+A_3B_3A_{1+2}
\nonumber \\
-2B_1V_{{\bf k}_1,{\bf k}_2,{\bf k}_3,{\bf k}_4}\Big \}\xi_{{\bf
k}_1}\xi_{{\bf k}_2}\xi_{{\bf k}_3}\xi_{{\bf k}_4}\nonumber \\
\times \delta({\bf k}_1+{\bf k}_2+{\bf k}_3+{\bf k}_4)d{\bf
k}_1d{\bf k}_2d{\bf k}_3d{\bf k}_4,
\nonumber \\
 \quad B_j\equiv
B_{k_j}, \ B_{j+l}\equiv B_{{\bf k}_j+{\bf k}_l}.
\end{eqnarray}

Canonical transform conserves the Hamiltonian structure so the
dynamical equations in new variables $R, \,\xi$ are given by:
\begin{equation}\label{hameqrxi1}
\frac{\partial \xi}{\partial t}=\frac{\delta H}{\delta R}, \quad
\frac{\partial R}{\partial t}=-\frac{\delta H}{\delta \xi}.
\end{equation}

\section{From complex to real Hopf equation}

We choose the cubic term of the generating functional $S$ in such
a way to remove linear instability at leading order. Similar to
Eqs. $(\ref{etaeq0}),(\ref{psieq0})$, we get from Eqs.
$(\ref{hamxi0}),(\ref{hamxi1}),(\ref{hameqrxi1})$ at leading
order of small parameter $|\nabla \xi|:$
\begin{subequations}
\begin{eqnarray}\label{xieq0}
\frac{\partial\xi}{\partial t}=\hat A R,
\\
\label{Req0} \frac{\partial R}{\partial t}=-\frac{1}{2}\big
(\nabla  R\big )^2.
\end{eqnarray}
\end{subequations}
Thus instead of the complex Hopf Eq. $(\ref{hopfp})$ (or Eq.
$(\ref{psieq0})$) we got the real Burgers Eq. $(\ref{Req0})$ for
new canonical variable $R$. It is important that the real Burgers
Eq. is well-posed.

Additional advantage of Eq. $(\ref{Req0})$ is that it can be
integrated by the the method of characteristics not only in two
dimensions as Eq. $(\ref{hopfp})$ but for three dimensional flow
also.

\section{Removal of instability from fourth order term}

Next step is to remove instability from the fourth order terms in
the Hamiltonian $(\ref{hamxi2})$ by a proper choice of matrix
element $V$.  We can take $V_{{\bf k}_1,{\bf k}_2,{\bf k}_3,{\bf
k}_4}$ in the following form:
\begin{equation}\label{vdef}
V_{{\bf k}_1,{\bf k}_2,{\bf k}_3,{\bf k}_4}=\alpha_1
k_1^2+\alpha_2 A_1(A_{2+3}+A_{2+4}+A_{3+4}),
\end{equation}
where $\alpha_1, \,\alpha_2$ are the real constants. The Eqs.
$(\ref{hamxi0}),(\ref{hamxi1}),(\ref{hamxi2})$ take the following
form:
\begin{eqnarray}
\label{hamxi0a} H_0=\frac{1}{2}\int \Big \{A_k |R_{\bf k}|^2+B_k
|\xi_{\bf k}|^2 \Big
\}d{\bf k}, \\
\label{hamxi1a} H_1=\frac{1}{4\pi}\int \Big[-({\bf k}_1\cdot{\bf
k}_2)R_{{\bf k}_1}R_{{\bf
k}_2}-\frac{1}{6}(A_1B_1+A_2B_2+A_3B_3)\xi_{{\bf k}_1}\xi_{{\bf
k}_2}\Big ] \nonumber\\
\times\xi_{{\bf k}_3}\delta({\bf k}_1+{\bf
k}_2+{\bf k}_3)d{\bf k}_1d{\bf k}_2d{\bf k}_3, \\
H_2=\frac{1}{8(2\pi)^2}\int \Big \{ ({\bf k}_1\cdot{\bf
k}_2)(A_{1+2}-A_1-A_2)
 \nonumber \\
+(3\alpha_1-1)(k_1^2A_2+k_2^2A_1)
 \nonumber \\
+\big [ \frac{1}{4}+3\alpha_2\big ]
A_1A_2[A_{1+3}+A_{2+3}+A_{1+4}+A_{2+4}]
 \nonumber \\
     +3\alpha_2A_1A_2(A_{1+2}+A_{3+4})
 \Big \}
\nonumber\\ \times
 R_{{\bf k}_1}R_{{\bf k}_2}\xi_{{\bf
k}_3}\xi_{{\bf k}_4}\delta({\bf k}_1+{\bf k}_2+{\bf k}_3+{\bf
k}_4)d{\bf k}_1d{\bf k}_2d{\bf k}_3d{\bf k}_4
\nonumber \\
\label{hamxi2a} +\frac{1}{8(2\pi)^2}\int \Big \{ -\sigma({\bf
k}_1\cdot{\bf k}_2)({\bf k}_3\cdot{\bf
k}_4)+\frac{1}{4}A_{1+2}^2B_{1+2}+A_3B_3A_{1+2}
\nonumber \\
-2B_1 \big [\alpha_1 k_1^2+\alpha_2 A_1(A_{2+3}+A_{2+4}+A_{3+4})
\big ]\Big \}\xi_{{\bf
k}_1}\xi_{{\bf k}_2}\xi_{{\bf k}_3}\xi_{{\bf k}_4}\nonumber \\
\times \delta({\bf k}_1+{\bf k}_2+{\bf k}_3+{\bf k}_4)d{\bf
k}_1d{\bf k}_2d{\bf k}_3d{\bf k}_4,
\nonumber \\
 \quad B_j\equiv
B_{k_j}, \ B_{j+l}\equiv B_{{\bf k}_j+{\bf k}_l}.
\end{eqnarray}

The dynamical equations, as follows from
$(\ref{hameqrxi1}),(\ref{hamxi0a})-(\ref{hamxi2a})$, are
\begin{eqnarray}\label{dynamicsRxi1a}
\frac{\partial R}{\partial t}=-\hat B \xi-\frac{1}{2}\big
(\nabla  R\big )^2+\frac{1}{2}\xi\hat A\hat B\xi+\frac{1}{4}\hat
A\hat B\xi^2+\frac{1}{4}\xi\hat A\big (\nabla R\big
)^2\nonumber \\
-\frac{1}{2}\xi\nabla R\cdot \nabla \hat A
R-\frac{1}{2}(1-3\alpha_1)\xi\big (\triangle R\big )\hat A R-\big
( \frac{1}{4}+3\alpha_2\big )\big (\hat A R\big )\hat A \big(\xi
\hat A R\big) \nonumber \\
-\frac{3\alpha_2}{2}\xi\hat A\big[\big (\hat A R\big)^2\big ]
-\frac{\sigma}{2}\nabla\cdot\Big [\nabla \xi(\nabla
\xi\cdot\nabla \xi)\Big ] -\frac{1}{8}\xi\hat A^2\big (\hat
B\xi^2\big ) \nonumber\\-\frac{1}{8}\big (\hat A\hat B\xi \big
)\hat A\xi^2 -\frac{1}{8}\hat A \hat B \big (\xi\hat
A\xi^2\big)-\frac{1-6\alpha_2}{4}\xi\hat A\big (\xi \hat A\hat
B\xi\big )-\frac{3}{4}\alpha_1\xi^2\triangle \hat B\xi
\nonumber\\+\frac{3}{4}\alpha_2\big (\hat A\hat B\xi\big )\hat
A\xi^2 -\frac{\alpha_1}{4}\triangle\hat B
\xi^3+\frac{3\alpha_2}{4}\hat A\hat B\big (\xi\hat A\xi^2\big ),
\\ \label{dynamicsRxi1b}
\frac{\partial \xi}{\partial t}=\hat A R-\nabla \cdot \big [
(\nabla R)\xi\big ]+\frac{1}{4}\nabla \cdot\Big[ \big (\nabla
R\big )\hat A\xi^2\Big ]-\frac{1}{4}\nabla \cdot \hat A\big
(\xi^2\nabla R\big )\nonumber\\-\frac{1}{4}\nabla\cdot \big
(\xi^2\hat A\nabla R\big
)+\frac{1-3\alpha_1}{4}\triangle\big(\xi^2 \hat AR\big
)+\frac{1-3\alpha_1}{4}\hat A\big (\xi^2\triangle R\big
)\nonumber\\+\big (\frac{1}{4}+3\alpha_2\big )\hat A\Big [\xi\hat
A\big (\xi\hat A R\big)\Big] +\frac{3}{2}\alpha_2\hat A\Big [\big
(\hat A\xi^2\big )\big (\hat A R\big)\Big ],
\end{eqnarray}
where $\hat B\equiv g-\sigma\triangle$, $\triangle\equiv
\frac{\partial ^2}{\partial x^2}+\frac{\partial ^2}{\partial
y^2}.$

To study linear stability of the Hamiltonian system in new
variable in respect to short wavelength perturbations one can set,
similar to Eq. $(\ref{etapsienvelope1})$, variables $\xi,\, R$ in
the following form
\begin{eqnarray}\label{xiRenvelope1}
\xi_{\bf k}=\xi_{0 \, {\bf k}}+\delta \xi_{\bf k}, \nonumber\\
R_{\bf k}=R_{0 \, {\bf k}}+\delta R_{\bf k},
\end{eqnarray}
with an assumption of an exponential dependence on time
\begin{eqnarray}\label{Rxiink3}
\delta\xi_{\bf k} \sim  e^{\nu_{\bf k} t}, \ \delta  R_{\bf
k}\sim  e^{\nu_{\bf k} t}.
\end{eqnarray}
Here $\xi_{0 \, {\bf k}},\, R_{0 \, {\bf k}}$ are solutions of
Eqs. $(\ref{dynamicsRxi1a}),(\ref{dynamicsRxi1b})$ and $\delta
\xi_{ {\bf k}},\, \delta R_{{\bf k}}$ are short wavelength
perturbations localized around wave vector ${\bf k}, |{\bf k}
|\gg q$, $q$ is a typical wavenumber for $\xi_{0 \, {\bf k}},\,
R_{0 \, {\bf k}}$.

We get, similar to Eqs.
$(\ref{H2lin}),(\ref{H2linEq}),(\ref{nuk2})$, for the perturbed
Hamiltonian,
\begin{eqnarray}\label{H2linxiR}
\delta H_0=\frac{1}{2} \int \tilde A_{\bf k}|\delta R_{\bf
k}|^2d{\bf k}+\frac{1}{2} \int \tilde B_{\bf k}|\delta\xi_{\bf
k}|^2 d{\bf k}+ \int (F_{\bf k}+iG_{\bf k})\delta R_{\bf k}
\delta\xi_{-{\bf k}} d{\bf k},
\nonumber\\
\tilde A_{\bf k}=\tilde A_{-{\bf k}},\ \tilde B_{\bf k}=\tilde
B_{-{\bf k}}, \ F_{\bf k}=F_{-{\bf k}}, \ G_{\bf k}=-G_{-{\bf k}},
\end{eqnarray}
the following expressions:
\begin{eqnarray}\label{AB3}
\tilde A_{\bf k}=A_k+k^2\xi_0+A_k\big [\frac{3}{2}\alpha_1
k^2+(\frac{1}{4}+3\alpha_2)A_k^2\big ]\xi_0^2+O(k^2\Theta\xi_0),
\nonumber\\
\tilde B_{\bf k}=B_k-A_kB_k\xi_0+ (\frac{1}{4}+3\alpha_2)A_k(\hat
A R_0)^2\nonumber \\
+ 3B_k\big [(\frac{1}{4}-\alpha_2)A_k^2 - \frac{1}{2}\alpha_1
k^2)\big ]\xi_0^2
 \nonumber \\
 +
O(g k \Theta\xi_0)+O(\sigma k^3\Theta\xi_0)+O( k^{0}v_0^2l_0^{-1}), \nonumber \\
F_{\bf k}=\big
[-\frac{1}{2}k^2(1-3\alpha_1)+(\frac{1}{4}+6\alpha_2)A_k^2 \big ]
 (\hat A R_0)\xi_0+O(kv_0\Theta),\nonumber\\
G_{\bf k}={\bf k}\cdot\nabla R_0+O(kv_0\Theta),
\end{eqnarray}
where $\xi_0=\frac{1}{2\pi}\int \xi_{\bf k} d{\bf k}, \
R_0=\frac{1}{2\pi}\int R_{\bf k} d{\bf k},$ and $\Theta\sim
|\nabla \xi_0|$ is the steepness. Similar to Eq. $(\ref{AB2})$, we
introduced here the typical value of fluid velocity,
$v_0\sim|\nabla R_0|$ and the typical scale, $l_0$ of variation
of $v_0$ and $\xi_0:$ $\Theta\sim \xi_0/l_0, \ |\nabla v_0|\sim
v_0/l_0.$

Eqs. $(\ref{nuk2}),(\ref{AB3})$ give instability growth rate. Our
purpose is to make these Eqs. well-posed for zero capillarity so
that we assume $\sigma=0$ and consider limit $k\to \infty$ which
means that $A_k=k$. It is convenient to rewrite Eqs.
$(\ref{nuk2}),(\ref{AB3})$ in dimensionless form as follows:
\begin{eqnarray}\label{AB3dimensionless}
 \tilde{\tilde A}_{\bf
k}=1+\lambda+\big [ \frac{3}{2}\alpha_1+\frac{1}{4}+3\alpha_2\big
] \lambda^2+O(\lambda\Theta),
\nonumber\\
\tilde{\tilde B}_{\bf
k}=1-\lambda+(\frac{1}{4}+3\alpha_2)\rho+3\big [
\frac{1}{4}-\alpha_2-\frac{\alpha_1}{2}\big ]
\lambda^2+O(\lambda\Theta)+O(\lambda^{-1}\rho\Theta),\nonumber \\
\tilde {F}_{\bf k}^2=\big [-
\frac{1}{4}+\frac{3}{2}\alpha_1+6\alpha_2\big
]^2\lambda^2\rho+O(\lambda\rho\Theta),
\nonumber \\
\tilde \nu_{\bf k}^2=\tilde{F}_{\bf k}^2-\tilde{\tilde A}_{\bf
k}\tilde{\tilde B}_{\bf k},
\end{eqnarray}
where $ \tilde{\tilde A}_{\bf k}=\tilde A_{\bf k}/k,$ $
\tilde{\tilde B}_{\bf k}=\tilde B_{\bf k}/g$, $\tilde {F}_{\bf
k}^2={F}_{\bf k}^2/(gk)$, $\tilde \nu_{\bf k}^2=(\nu_{\bf
k}+iG_{\bf k})^2/(gk)$. The system $(\ref{AB3dimensionless})$ is
described by the two independent dimensionless parameters
$\lambda\equiv k\xi_0$ and $\rho\equiv k(\hat A R_0)^2/g$ which
reflects the freedom of choice of an initial surface elevation
and an initial velocity. Condition of applicability of Eqs.
$(\ref{AB3})$ is $kl_0\ll 1$, which gives $|\lambda| \gg  \Theta$
in dimensionless variables $\rho,\, \lambda.$ Parameter $\rho$
can take any nonnegative value because it depends on the fluid
velocity which can be arbitrary. We want to choose $\alpha_1$ and
$\alpha_2$ to ensure that Eqs.
$(\ref{hameqrxi1}),(\ref{hamxi0a}),(\ref{hamxi1a}),(\ref{hamxi2a})$
are well-posed and stable, which means that $\tilde \nu_{\bf k}^2<
0$, for any value of $\lambda$ and any nonnegative value of
$\rho$.

First step is to analyze the system $(\ref{AB3dimensionless})$ in
the limit $\Theta\to 0$ which means that we first neglect
$O(\ldots\Theta)$ terms in $(\ref{AB3dimensionless})$. Assume that
$g\neq 0$ then the necessary condition for $\tilde \nu_{\bf k}^2<
0$ is to have $\tilde{\tilde A}_{\bf k}\tilde{\tilde B}_{\bf
k}>0$ which means that either $\tilde{\tilde A}_{\bf k}>0$ and
$\tilde{\tilde B}_{\bf k}>0,$ or $\tilde{\tilde A}_{\bf k}<0$ and
$\tilde{\tilde B}_{\bf k}<0$. It easy to show that the second case
can not be realized for Eqs. $(\ref{AB3dimensionless})$ so we
consider the first case of positive $\tilde{\tilde A}_{\bf k}$
and $\tilde{\tilde B}_{\bf k}.$ Inequality $\tilde{\tilde A}_{\bf
k}>0$ gives $\beta_1>-1/3$ and inequality $\tilde{\tilde B}_{\bf
k}|_{\rho=0}>0$ gives $\beta_1 <1/6$ which together result in
\begin{eqnarray}\label{beta1cond}
-\frac{\sqrt{3}}{6}< \beta_1< \frac{\sqrt{3}}{12}, \quad
\beta_1=\frac{\sqrt{3}}{2}(\frac{3}{2}\alpha_1+3\alpha_2-\frac{1}{3}).
\end{eqnarray}

Provided $(\ref{beta1cond})$ is satisfied, the sufficient
condition for absence of instability, $\tilde \nu_{\bf k}^2< 0$,
is to have term $\propto \rho$ in $\tilde \nu_{\bf k}^2$ to be
negative for any $\lambda$, which means that
\begin{eqnarray}\label{intermcond}
\big [- \frac{1}{4}+\frac{3}{2}\alpha_1+6\alpha_2\big
]^2\lambda^2-(\frac{1}{4}+3\alpha_2)\tilde{\tilde A}_{\bf k}<0.
\end{eqnarray}
This inequality is satisfied for any $\lambda$ provided
\begin{eqnarray}\label{beta2cond}
\beta_1^2+\beta_2^2<\frac{1}{12}, \quad
\beta_2=\frac{1}{4}(-1+3\alpha_1+18\alpha_2).
\end{eqnarray}
Thus $\tilde \nu_{\bf k}^2< 0$ for any $\lambda$ and $\rho\ge 0$
provided inequalities $(\ref{beta1cond})$ and $(\ref{beta2cond})$
hold. It corresponds in $(\alpha_1,\,\alpha_2)$ plane to the inner
part of the ellipse defined by $(\ref{beta2cond})$ and bounded by
two parallel lines defined by $(\ref{beta1cond})$ (see the filled
area in Figure 1). The center of the ellipse is located at
$\alpha_1=\frac{1}{6}, \quad \alpha_2=\frac{1}{36}$ (point $A$ in
Figure 1). So the choice of $\alpha_1$ and $\alpha_2$ is not
unique for $g\neq 0$ in the limit $\Theta\to 0$ and is determined
by $(\ref{beta1cond}),(\ref{beta2cond})$.

For $g=0$ Eqs. $(\ref{AB3dimensionless})$ are reduced to
\begin{eqnarray}\label{AB3dimensionlessg0}
 \tilde{\tilde A}_{\bf
k}=1+\lambda+\big [ \frac{3}{2}\alpha_1+\frac{1}{4}+3\alpha_2\big
] \lambda^2+O(\lambda\Theta),
\nonumber \\
\tilde{\tilde \nu}_{\bf k}^2=\big [-
\frac{1}{4}+\frac{3}{2}\alpha_1+6\alpha_2\big ]^2\lambda^2
%\nonumber \\
%\qquad \qquad \qquad \qquad
-\big[\frac{1}{4}+3\alpha_2+O(\lambda^{-1}\Theta)\big
]\tilde{\tilde A}_{\bf k}+O(\lambda\Theta), \nonumber \\
\tilde{\tilde \nu}_{\bf k}^2= \frac{\xi_0^2}{\lambda^2(\hat A
R_0)^2}(\nu_{\bf k}+iG_{\bf k})^2.
\end{eqnarray}
It  follows from  Eq. $(\ref{AB3dimensionlessg0})$ that
$\tilde{\tilde \nu}_{\bf k}^2< 0$ in the limit $\Theta\to 0$
provided condition $(\ref{beta2cond})$ is satisfied, which
corresponds in $(\alpha_1,\,\alpha_2)$ plane to the inner part of
the ellipse defined by $(\ref{beta2cond})$ (see Figure 2) in
contrast with the case of nonzero gravity.

For $k\to \infty$ we get from $(\ref{AB3})$ for $g\neq 0$:
\begin{eqnarray}\label{nuka1a2}
\nu_{\bf k}=-i{\bf k}\cdot\nabla R_0\pm ik^{5/2}g^{1/2}3^{1/2}
\big [ \frac{3}{2}\alpha_1+\frac{1}{4}+3\alpha_2\big
]^{1/2}\nonumber \\
\times\big [ \frac{1}{4}-\alpha_2-\frac{\alpha_1}{2}\big
]^{1/2}\xi_0^2
\end{eqnarray}
and
\begin{eqnarray}\label{nuka1a2g0}
\nu_{\bf k}=-i{\bf k}\cdot\nabla R_0\pm ik^2  2^{-3/2}3 \nonumber \\
\times\big [
-2\alpha_1^2+4\alpha_2(1-6\alpha_2)+\alpha_1(1-12\alpha_2)]^{1/2}\xi_0\hat
A R_0
\end{eqnarray}
for $g=0$.

Note that for parameters satisfying inequalities
$(\ref{beta1cond}),(\ref{beta2cond})$, the real part of $\nu_{\bf
k}$ is zero even for $kh\sim 1$, where $\nu_{\bf k}$ takes the
following form:
\begin{eqnarray}\label{nukAa1a2}
\nu_{\bf k}=-i{\bf k}\cdot\nabla R_0\pm iA_k^{1/2}\big [
g+(\frac{1}{4}+3\alpha_2)\, A_k(\hat A R_0)^2\big ]^{1/2},
\end{eqnarray}
so that in new canonical variables
$(\ref{vdef}),(\ref{beta1cond}),(\ref{beta2cond})$ the
instability is absent even for intermediate values of $k\sim
1/h.$ Another remark is that these variables leave problem
well-posed for $\sigma\neq 0$ also but that case is not so
interesting because Eqs.
$(\ref{hameqrxi1}),(\ref{hamxi0a}),(\ref{hamxi1a}),(\ref{hamxi2a})$
well-posed even in original variables $\eta, \ \psi$ for
$\sigma\neq 0$.

Now we make the second step and assume that $\Theta$ is small but
nonzero in Eqs. $(\ref{AB3dimensionless})$ and
$(\ref{AB3dimensionlessg0})$. Terms $O(\ldots\Theta)$ in Eqs.
$(\ref{AB3dimensionless})$ and $(\ref{AB3dimensionlessg0})$ are
not sign-definite and their values depends on horizontal
coordinates $x, \,y$ and time according to dynamical Eqs.
$(\ref{hameqrxi1}),(\ref{hamxi0a}),(\ref{hamxi1a}),(\ref{hamxi2a})$.
Generally these terms result in shrinking of the area of
stability, $\tilde \nu_{\bf k}^2< 0$ in $(\alpha_1,\,\alpha_2)$
plane. Figures 3 and 4 show shrinking of the stability area for
the particular choice of terms $O(\ldots\Theta)$ for $g\neq 0$.
Each curve in Figure 4 corresponds to the stability boundary,
$\max\limits_{\lambda, \, \rho}\tilde \nu_{\bf k}^2=0,$ for the
particular value of $\Theta.$ The additional requirement is that
\begin{eqnarray}\label{beta1condTheta}
\frac{b_1}{12}\Theta(2+b_1\Theta)<\alpha_2+\frac{\alpha_1}{2}<\frac{1}{6}+\frac{b_2}{12}\Theta(2-b_2\Theta),
\end{eqnarray}
which is a generalization of $(\ref{beta1cond})$ for nonzero
$\Theta$. Here we assume $O(\lambda\Theta)=b_1\lambda\Theta$ and
$O(\lambda\Theta)=b_2\lambda\Theta$ in rhs of Eqs.
$(\ref{AB3dimensionless})$ for $\tilde{\tilde A}_{\bf k}$ and
$\tilde{\tilde B}_{\bf k}$, respectively. Calculating  curves in
Figures 3 and 4 we set $b_1=b_2=-1$ as a typical example.
Condition $(\ref{beta1condTheta})$ result in additional cutting
of curves $\max\limits_{\lambda, \, \rho}\tilde \nu_{\bf k}^2=0$
in Figures 3 and 4 for $\Theta=0, \, 0.01, \,0.02, \,0.025 $ and $
0.05.$
 The system $(\ref{AB3dimensionless})$ is stable inside each
curve in Figure 4 for given $\Theta.$ For $\Theta<0.1$, the width
of region between solid curve ($\Theta=0$) and curves with
$\Theta\neq 0$ scales approximately as $\Theta$. For $\Theta>0.1$
the region of stability quickly shrinks to zero as $\Theta$
approaches $\simeq 0.1389$. Note that these numerical values are
non-universal and depend on the numerical coefficient in
$O(\ldots\Theta)$ terms. In a similar way, Figures 5,6 show
shrinking of the stability area for zero gravity case.

Our objective is to find parameters $\alpha_1, \, \alpha_2$
corresponding to stability, $\tilde \nu_{\bf k}^2< 0$, with the
largest possible $\Theta$.   For the case $g\neq 0$ this is
achieved if the maximum of $\tilde \nu_{\bf k}^2|_{\Theta=0}$, as
a function of $\lambda, \, \rho$, is not only negative but
minimum as a function of $\alpha_1$ and $\alpha_2$, i.e. we want
to find $\min\limits_{\alpha_1, \,\alpha_2} \max\limits_{\lambda,
\, \rho}\tilde \nu_{\bf k}^2\big|_{\Theta=0}$. This ensures that
the system $(\ref{AB3dimensionless})$ is the most stable system
for $\Theta \to 0$ or, in other words, the system is the most
rigid one.  Because $\rho\ge 0$ we have to set $\rho=0$ to find
$\max\limits_{\lambda, \, \rho}\tilde \nu_{\bf k}^2$. Then we
obtain that $\min\limits_{\alpha_1, \,\alpha_2}
\max\limits_{\lambda, \, \rho}\tilde \nu_{\bf k}^2=\tilde
\nu_{\bf k}^2\big|_{\rho=0, \, \lambda=0}=-1$. This minimum is
attained provided $3\alpha_1+6\alpha_2=-1/2$ and
$\frac{1}{6}(1-\sqrt{5})<\alpha_1<1/3$. We also want  to have the
most stable system for $k\to\infty$ ($\rho\to \infty$ and
$\lambda\to \infty$). This is achieved provided  the coefficients
in front of the leading order terms $\lambda^4$ and $\lambda^2
\rho$ in $\tilde \nu_{\bf k}^2$ are minimums. The coefficient for
$\lambda^4$ is already minimum from condition
$3\alpha_1+6\alpha_2=1/2$, while the coefficient for $\lambda^2
\rho$ is $-3/16+9\alpha_1^2/4$, i.e. we have minimum for
\begin{eqnarray}\label{alpha1alpha2p1p6}
\alpha_1=0, \quad \alpha_2=1/12, \quad g\neq 0,
\end{eqnarray}
which corresponds to point $B$ in Figure 1. This choice of
parameters $\alpha_1$ and $\alpha_2$ is optimal to keep the
system $(\ref{AB3dimensionless})$ stable for the largest possible
$\Theta$, i.e. for the largest possible nonlinearity.

In a similar way, for zero gravity, $g=0,$ the system
$(\ref{AB3dimensionlessg0})$ is the most stable provided we find
$\alpha_1, \,\alpha_2$ which correspond to
$\min\limits_{\alpha_1, \,\alpha_2}
\max\limits_{\lambda}\tilde{\tilde \nu}_{\bf k}^2\big|_{\Theta=0}$
for Eqs. $(\ref{AB3dimensionlessg0})$. Maximum
$\max\limits_{\lambda}\tilde{\tilde \nu}_{\bf
k}^2\big|_{\Theta=0}$ is attained for
$$\lambda=\frac{(1+12\alpha_2)}{9}\big
[2\alpha_1^2+4\alpha_2(6\alpha_2-1)+\alpha_1(12\alpha_2-1)\big
]^{-1},$$ and $\min\limits_{\alpha_1, \,\alpha_2}
\max\limits_{\lambda}\tilde{\tilde \nu}_{\bf
k}^2\big|_{\Theta=0}=\frac{2}{3}(-2+\sqrt{3})$ is attained
provided
\begin{eqnarray}\label{alpha1alpha2p1p6g0}
\alpha_1=-\frac{1}{2}+3^{-1/2}, \quad
\alpha_2=\frac{1}{4}-3^{3/2}, \quad g=0,
\end{eqnarray}
which corresponds to point $E$ in Figure 2. This choice of
parameters $\alpha_1$ and $\alpha_2$ is optimal to keep the
system $(\ref{AB3dimensionless})$ stable for the largest possible
$\Theta$, i.e. for the largest possible nonlinearity.

Thus we can choose $\alpha_1, \, \alpha_2$ from the conditions
$(\ref{beta1cond})$ and $(\ref{beta2cond})$ to make Eqs.
$(\ref{hameqrxi1}),(\ref{hamxi0a})-(\ref{hamxi2a})$ (or,
equivalently, Eqs. $(\ref{dynamicsRxi1a}),(\ref{dynamicsRxi1b})$)
well-posed for any value of $\sigma, \, g$ and arbitrary depth of
fluid. To find dynamics of free surface, one can solve Eqs. for
$R, \, \xi$ using Eqs.
$(\ref{hameqrxi1}),(\ref{hamxi0a})-(\ref{hamxi2a})$ and
conditions $(\ref{beta1cond}),(\ref{beta2cond}).$ This is the
main result of  this Article. To recover physical variables
$\Psi, \, \eta$ from given $R, \, \xi$ one can use Eqs.
$(\ref{etaeq1}),(\ref{psieq1}),(\ref{vdef}).$

Now we can return to the comment in Section 5 about interpretation
of ill-posedness of Eqs.
$(\ref{hamfoureq1}),(\ref{ham0})-(\ref{ham2})$  as violation of
perturbation expansion $(\ref{hamserdef})$ for $k\eta_0\gtrsim 1.$
For the new canonical variables $\xi, \, R$ perturbation expansion
is still formally violated for $k\xi_0\gtrsim 1$ because
contribution from the quadratic Hamiltonian
$(\ref{H2linxiR}),(\ref{AB3})$ is not small compare with other
terms in the Hamiltonian $(\ref{hamxi0a})-(\ref{hamxi2a})$.
However this violation does cause any problem because there is no
short wavelength instability in the new canonical variables and
the system $(\ref{hameqrxi1}),(\ref{hamxi0a})-(\ref{hamxi2a})$ is
well-posed. In other words,  the new canonical variables $\xi, \,
R$ provide purely physical way to regularize short wavelengths
without introduction any artificial viscosity.

As follows from Eqs. $(\ref{beta1cond}),(\ref{beta2cond}),$ the
new canonical variables $\xi, \, R$ are not uniquely determined
from the condition of well-posedness of the dynamical Eqs.
$(\ref{hameqrxi1}),(\ref{hamxi0a})-(\ref{hamxi2a})$ because
parameters $(\alpha_1,\,\alpha_2)$ can take any valued from
filled area in Figures 1 and 2. However the choice of
$(\alpha_1,\,\alpha_2)$ is unique provided we additionally
require the system
$(\ref{hameqrxi1}),(\ref{hamxi0a})-(\ref{hamxi2a})$ to be free of
short wavelength instability for the largest possible slopes
$\Theta$, i.e. for the largest possible nonlinearity. This gives
the conditions $(\ref{alpha1alpha2p1p6})$ for $g\neq 0$ and
$(\ref{alpha1alpha2p1p6g0})$ for $g=0.$ We refer to the variables
$\xi, \, R$, defined in Eqs.
$(\ref{etaeq1}),(\ref{psieq1}),(\ref{vdef}),(\ref{alpha1alpha2p1p6})$
and $(\ref{alpha1alpha2p1p6g0})$, as {\it the optimal canonical
variables}. For some extent similar results were obtained by
Dyachenko \cite{Dyachenko2004} for particular case of
two-dimensional flow. We conjecture that the optimal canonical
variables, which allow well-posedness of the dynamical Eqs., exist
in all orders of nonlinearity. However additional research
necessary to decide if the optimal canonical variables exist and
unique in higher (fifth etc.) order of nonlinearity. We also
conjecture that the optimal canonical variables $\xi, \, R$  would
allow simulation with higher steepness compare  with standard
variables $\Psi, \, \eta.$

\section{Special cases}

There are a number of important special cases of the optimal
canonical variables.  Here we use $(\ref{alpha1alpha2p1p6})$ and
$(\ref{alpha1alpha2p1p6g0})$. We give here expression for the
Hamiltonian only. The dynamical Eqs. can be obtained either from
$(\ref{hameqrxi1})$ or directly from Eqs.
$(\ref{dynamicsRxi1a}),(\ref{dynamicsRxi1b})$.

\subsection{Deep water limit}
For $g\neq 0, \ h\to \infty$, $A_k= k$ and  Eqs.
$(\ref{hamxi0a})-(\ref{hamxi2a})$ take the form
\begin{eqnarray}
\label{hamxiinfinite0} H_0=\frac{1}{2}\int \Big \{k |R_{\bf
k}|^2+B_k |\xi_{\bf k}|^2 \Big
\}d{\bf k}, \quad B_k=g+\sigma k^2, \\
\label{hamxiinfinite1} H_1=\frac{1}{4\pi}\int \Big[-({\bf
k}_1\cdot{\bf k}_2)R_{{\bf k}_1}R_{{\bf
k}_2}-\frac{1}{6}(k_1B_1+k_2B_2+k_3B_3)\xi_{{\bf k}_1}\xi_{{\bf
k}_2}\Big ] \nonumber\\
\times\xi_{{\bf k}_3}\delta({\bf k}_1+{\bf
k}_2+{\bf k}_3)d{\bf k}_1d{\bf k}_2d{\bf k}_3, \\
H_2=\frac{1}{8(2\pi)^2}\int \Big \{ ({\bf k}_1\cdot{\bf
k}_2)(|{\bf k}_1+{\bf
k}_2|-k_1-k_2)
\nonumber\\
-(k_1^2k_2+k_2^2k_1)\nonumber
\\+\frac{1}{2}k_1k_2\big [|{\bf
k}_1+{\bf k}_3|+|{\bf k}_2+{\bf k}_3|+|{\bf k}_1+{\bf k}_4|+|{\bf
k}_2+{\bf k}_4|\big ]
\nonumber\\
+\frac{1}{4}k_1k_2\big [|{\bf k}_1+{\bf k}_2|+|{\bf k}_3+{\bf
k}_4|\big ]\Big \}\nonumber
\\
 \times R_{{\bf k}_1}R_{{\bf
k}_2}\xi_{{\bf k}_3}\xi_{{\bf k}_4}\delta({\bf k}_1+{\bf
k}_2+{\bf k}_3+{\bf k}_4)d{\bf k}_1d{\bf k}_2d{\bf k}_3d{\bf
k}_4 \nonumber \\
\label{hamxiinfinite2} +\frac{1}{8(2\pi)^2}\int \Big \{
-\sigma({\bf k}_1\cdot{\bf k}_2)({\bf k}_3\cdot{\bf
k}_4)+\frac{1}{4}|{\bf k}_1+{\bf k}_2|^2B_{1+2}+k_3B_3|{\bf
k}_1+{\bf k}_2|\nonumber \\
-\frac{1}{6}B_1 k_1(|{\bf k}_2+{\bf k}_3|+|{\bf k}_2+{\bf
k}_4|+|{\bf k}_3+{\bf k}_4|) \Big \}
\nonumber\\
\times
 \xi_{{\bf k}_1}\xi_{{\bf k}_2}\xi_{{\bf k}_3}\xi_{{\bf
k}_4}
 \delta({\bf k}_1+{\bf k}_2+{\bf k}_3+{\bf k}_4)d{\bf
k}_1d{\bf k}_2d{\bf k}_3d{\bf k}_4.
\end{eqnarray}

\subsubsection{Zero gravity and capillarity $g=\sigma=0$}

\begin{eqnarray}
\label{hamxiinfinite0a} H_0=\frac{1}{2}\int k |R_{\bf k}|^2 d{\bf k}, \\
\label{hamxiinfinite1a} H_1=-\frac{1}{4\pi}\int ({\bf
k}_1\cdot{\bf k}_2)R_{{\bf k}_1}R_{{\bf k}_2}\xi_{{\bf
k}_3}\delta({\bf k}_1+{\bf
k}_2+{\bf k}_3)d{\bf k}_1d{\bf k}_2d{\bf k}_3, \\
H_2=\frac{1}{8(2\pi)^2}\int \Big \{ ({\bf k}_1\cdot{\bf
k}_2)(|{\bf k}_1+{\bf k}_2|-k_1-k_2)
\nonumber\\
+(-\frac{5}{2}+\sqrt{3})(k_1^2k_2+k_2^2k_1)\nonumber
\\+(1-3^{5/2})k_1k_2\big [|{\bf k}_1+{\bf
k}_3|+|{\bf k}_2+{\bf k}_3|+|{\bf k}_1+{\bf k}_4|+|{\bf k}_2+{\bf
k}_4|\big ]
\nonumber\\
+(\frac{3}{4}-3^{5/2})k_1k_2\big [|{\bf k}_1+{\bf k}_2|+|{\bf
k}_3+{\bf k}_4|\big ]\Big \} \nonumber
\\
 \times R_{{\bf k}_1}R_{{\bf
k}_2}\xi_{{\bf k}_3}\xi_{{\bf k}_4}\delta({\bf k}_1+{\bf
k}_2+{\bf k}_3+{\bf k}_4)d{\bf k}_1d{\bf k}_2d{\bf k}_3d{\bf
k}_4.\label{hamxiinfinite2a}
\end{eqnarray}

\subsubsection{Zero gravity, $g=0,$ and nonzero capillarity  $\sigma\neq 0$}

\begin{eqnarray}
\label{hamxiinfinite0b} H_0=\frac{1}{2}\int \Big \{k |R_{\bf
k}|^2+\sigma k^2 |\xi_{\bf k}|^2 \Big
\}d{\bf k}, \\
\label{hamxiinfinite1b} H_1=\frac{1}{4\pi}\int \Big[-({\bf
k}_1\cdot{\bf k}_2)R_{{\bf k}_1}R_{{\bf
k}_2}-\frac{\sigma}{6}(k_1^3+k_2^3+k_3^3)\xi_{{\bf k}_1}\xi_{{\bf
k}_2}\Big ] \nonumber\\
\times\xi_{{\bf k}_3}\delta({\bf k}_1+{\bf
k}_2+{\bf k}_3)d{\bf k}_1d{\bf k}_2d{\bf k}_3, \\
H_2=\frac{1}{8(2\pi)^2}\int \Big \{ ({\bf k}_1\cdot{\bf
k}_2)(|{\bf k}_1+{\bf k}_2|-k_1-k_2)\nonumber
\\
+(-\frac{5}{2}+\sqrt{3})(k_1^2k_2+k_2^2k_1)\nonumber
\\+(1-3^{5/2})k_1k_2\big [|{\bf k}_1+{\bf
k}_3|+|{\bf k}_2+{\bf k}_3|+|{\bf k}_1+{\bf k}_4|+|{\bf k}_2+{\bf
k}_4|\big ]
\nonumber\\
+(\frac{3}{4}-3^{5/2})k_1k_2\big [|{\bf k}_1+{\bf k}_2|+|{\bf
k}_3+{\bf k}_4|\big ]\Big \} \nonumber
\\
 \times R_{{\bf k}_1}R_{{\bf
k}_2}\xi_{{\bf k}_3}\xi_{{\bf k}_4}\delta({\bf k}_1+{\bf
k}_2+{\bf k}_3+{\bf k}_4)d{\bf k}_1d{\bf k}_2d{\bf k}_3d{\bf
k}_4 \nonumber \\
\label{hamxiinfinite2b} +\frac{\sigma}{8(2\pi)^2}\int \Big \{
-({\bf k}_1\cdot{\bf k}_2)({\bf k}_3\cdot{\bf
k}_4)+\frac{1}{4}|{\bf k}_1+{\bf k}_2|^4+k_3^3|{\bf
k}_1+{\bf k}_2|\nonumber \\
-k_1^3 \big [(-1+3^{-1/2}) k_1+(\frac{1}{2}-3^{3/2}2)(|{\bf
k}_2+{\bf k}_3|+|{\bf k}_2+{\bf k}_4|+|{\bf k}_3+{\bf k}_4|)\big
] \Big \}
\nonumber \\
\times \xi_{{\bf k}_1}\xi_{{\bf k}_2}\xi_{{\bf k}_3}\xi_{{\bf
k}_4}
 \delta({\bf k}_1+{\bf k}_2+{\bf k}_3+{\bf k}_4)d{\bf
k}_1d{\bf k}_2d{\bf k}_3d{\bf k}_4.
\end{eqnarray}

\subsubsection{Nonzero gravity, $g\neq 0,$ and zero capillarity  $\sigma=0$}

\begin{eqnarray}
\label{hamxiinfinite0c} H_0=\frac{1}{2}\int \Big \{k |R_{\bf
k}|^2+g |\xi_{\bf k}|^2 \Big
\}d{\bf k},  \\
\label{hamxiinfinite1c} H_1=\frac{1}{4\pi}\int \Big[-({\bf
k}_1\cdot{\bf k}_2)R_{{\bf k}_1}R_{{\bf
k}_2}-\frac{g}{6}(k_1+k_2+k_3)\xi_{{\bf k}_1}\xi_{{\bf
k}_2}\Big ] \nonumber\\
\times\xi_{{\bf k}_3}\delta({\bf k}_1+{\bf
k}_2+{\bf k}_3)d{\bf k}_1d{\bf k}_2d{\bf k}_3, \\
H_2=\frac{1}{8(2\pi)^2}\int \Big \{ ({\bf k}_1\cdot{\bf
k}_2)(|{\bf k}_1+{\bf k}_2|-k_1-k_2)-(k_1^2k_2+k_2^2k_1)\nonumber
\\+\frac{1}{2}k_1k_2\big [|{\bf k}_1+{\bf
k}_3|+|{\bf k}_2+{\bf k}_3|+|{\bf k}_1+{\bf k}_4|+|{\bf k}_2+{\bf
k}_4|\big ]
\nonumber\\
+\frac{1}{4}k_1k_2\big [|{\bf k}_1+{\bf k}_2|+|{\bf k}_3+{\bf
k}_4|\big ]\Big \} \nonumber
\\
 \times R_{{\bf k}_1}R_{{\bf
k}_2}\xi_{{\bf k}_3}\xi_{{\bf k}_4}\delta({\bf k}_1+{\bf
k}_2+{\bf k}_3+{\bf k}_4)d{\bf k}_1d{\bf k}_2d{\bf k}_3d{\bf
k}_4 \nonumber \\
\label{hamxiinfinite2c} +\frac{g}{8(2\pi)^2}\int \Big \{
\frac{1}{4}|{\bf k}_1+{\bf k}_2|^2+k_3|{\bf
k}_1+{\bf k}_2|\nonumber \\
-\frac{1}{6}k_1(|{\bf k}_2+{\bf k}_3|+|{\bf k}_2+{\bf k}_4|+|{\bf
k}_3+{\bf k}_4|) \Big \}
\nonumber \\
\times \xi_{{\bf k}_1}\xi_{{\bf k}_2}\xi_{{\bf k}_3}\xi_{{\bf
k}_4}
 \delta({\bf k}_1+{\bf k}_2+{\bf k}_3+{\bf k}_4)d{\bf
k}_1d{\bf k}_2d{\bf k}_3d{\bf k}_4.
\end{eqnarray}

\subsection{Shallow water limit}

Shallow water limit corresponds to $kh\to 0$. In that limit
$A_k\to k^2 h$. Eqs. $(\ref{hamxi0a})-(\ref{hamxi2a})$ take the
following form for $g\neq 0:$
\begin{eqnarray}
\label{hamxishallow0} H_0=\frac{1}{2}\int \Big \{k^2 h |R_{\bf
k}|^2+B_k |\xi_{\bf k}|^2 \Big
\}d{\bf k}, \quad B_k=g+\sigma k^2, \\
\label{hamxishallow1} H_1=\frac{1}{4\pi}\int \Big[-({\bf
k}_1\cdot{\bf k}_2)R_{{\bf k}_1}R_{{\bf
k}_2}-\frac{h}{6}(k_1^2B_1+k_2^2B_2+k_3^2B_3)\xi_{{\bf
k}_1}\xi_{{\bf
k}_2}\Big ] \nonumber\\
\times\xi_{{\bf k}_3}\delta({\bf k}_1+{\bf
k}_2+{\bf k}_3)d{\bf k}_1d{\bf k}_2d{\bf k}_3, \\
H_2=\frac{h}{8(2\pi)^2}\int \Big \{ 2({\bf k}_1\cdot{\bf k}_2)^2
-(k_1^2k_2^2+k_2^2k_1^2)\nonumber
\\+h^2\frac{1}{2}k_1^2k_2^2\big [|{\bf
k}_1+{\bf k}_3|^2+|{\bf k}_2+{\bf k}_3|^2+|{\bf k}_1+{\bf
k}_4|^2+|{\bf k}_2+{\bf k}_4|^2\big ]
\nonumber\\
+h^2 \frac{1}{4}k_1^2k_2^2\big [|{\bf k}_1+{\bf k}_2|^2+|{\bf
k}_3+{\bf k}_4|^2\big ]\Big \}\nonumber
\\
 \times R_{{\bf k}_1}R_{{\bf
k}_2}\xi_{{\bf k}_3}\xi_{{\bf k}_4}\delta({\bf k}_1+{\bf
k}_2+{\bf k}_3+{\bf k}_4)d{\bf k}_1d{\bf k}_2d{\bf k}_3d{\bf
k}_4 \nonumber \\
\label{hamxishallow2} +\frac{1}{8(2\pi)^2}\int \Big \{
-\sigma({\bf k}_1\cdot{\bf k}_2)({\bf k}_3\cdot{\bf k}_4)
\nonumber\\
+\frac{h^2}{4}|{\bf k}_1+{\bf k}_2|^4B_{1+2}+h^2k_3^2B_3|{\bf
k}_1+{\bf k}_2|^2\nonumber \\
-\frac{B_1k_1^2h^2}{6} (|{\bf k}_2+{\bf k}_3|^2+|{\bf k}_2+{\bf
k}_4|^2+|{\bf k}_3+{\bf k}_4|^2)\Big \}
\nonumber\\
\times
 \xi_{{\bf k}_1}\xi_{{\bf k}_2}\xi_{{\bf k}_3}\xi_{{\bf
k}_4}
 \delta({\bf k}_1+{\bf k}_2+{\bf k}_3+{\bf k}_4)d{\bf
k}_1d{\bf k}_2d{\bf k}_3d{\bf k}_4.
\end{eqnarray}

\subsubsection{Zero gravity and capillarity $g=\sigma=0$}

\begin{eqnarray}
\label{hamxishallow0a} H_0=\frac{1}{2}\int k^2 h |R_{\bf k}|^2 d{\bf k}, \\
\label{hamxishallow1a} H_1=-\frac{1}{4\pi}\int ({\bf
k}_1\cdot{\bf k}_2)R_{{\bf k}_1}R_{{\bf k}_2}\xi_{{\bf
k}_3}\delta({\bf k}_1+{\bf
k}_2+{\bf k}_3)d{\bf k}_1d{\bf k}_2d{\bf k}_3, \\
H_2=\frac{h}{8(2\pi)^2}\int \Big \{ 2({\bf k}_1\cdot{\bf
k}_2)^2+(-\frac{5}{2}+\sqrt{3})(k_1^2k_2^2+k_2^2k_1^2)
\nonumber\\
+h^2(1-3^{5/2})k_1^2k_2^2\big [|{\bf k}_1+{\bf k}_3|^2+|{\bf
k}_2+{\bf k}_3|^2+|{\bf k}_1+{\bf k}_4|^2+|{\bf k}_2+{\bf
k}_4|^2\big ]
\nonumber\\
+h^2 (\frac{3}{4}-3^{5/2})k_1^2k_2^2\big [|{\bf k}_1+{\bf
k}_2|^2+|{\bf k}_3+{\bf k}_4|^2\big ]\Big \} \nonumber
\\
 \times R_{{\bf k}_1}R_{{\bf
k}_2}\xi_{{\bf k}_3}\xi_{{\bf k}_4}\delta({\bf k}_1+{\bf
k}_2+{\bf k}_3+{\bf k}_4)d{\bf k}_1d{\bf k}_2d{\bf k}_3d{\bf
k}_4. \label{hamxishallow2a}
\end{eqnarray}

\subsubsection{Zero gravity, $g=0,$ and nonzero capillarity  $\sigma\neq 0$}

\begin{eqnarray}
\label{hamxishallow0b} H_0=\frac{1}{2}\int \Big \{k^2 h |R_{\bf
k}|^2+\sigma k^2 |\xi_{\bf k}|^2 \Big
\}d{\bf k}, \\
\label{hamxishallow1b} H_1=\frac{1}{4\pi}\int \Big[-({\bf
k}_1\cdot{\bf k}_2)R_{{\bf k}_1}R_{{\bf k}_2}-\frac{\sigma
h}{6}(k_1^4+k_2^4+k_3^4)\xi_{{\bf k}_1}\xi_{{\bf
k}_2}\Big ] \nonumber\\
\times\xi_{{\bf k}_3}\delta({\bf k}_1+{\bf
k}_2+{\bf k}_3)d{\bf k}_1d{\bf k}_2d{\bf k}_3, \\
H_2=\frac{h}{8(2\pi)^2}\int \Big \{ 2({\bf k}_1\cdot{\bf
k}_2)^2+(-\frac{5}{2}+\sqrt{3})(k_1^2k_2^2+k_2^2k_1^2)\nonumber
\\+h^2(1-3^{5/2})k_1^2k_2^2\big [|{\bf
k}_1+{\bf k}_3|^2+|{\bf k}_2+{\bf k}_3|^2+|{\bf k}_1+{\bf
k}_4|^2+|{\bf k}_2+{\bf k}_4|^2\big ]
\nonumber\\
+h^2 (\frac{3}{4}-3^{5/2})k_1^2k_2^2\big [|{\bf k}_1+{\bf
k}_2|^2+|{\bf k}_3+{\bf k}_4|^2\big ]\Big \} \nonumber
\\
 \times R_{{\bf k}_1}R_{{\bf
k}_2}\xi_{{\bf k}_3}\xi_{{\bf k}_4}\delta({\bf k}_1+{\bf
k}_2+{\bf k}_3+{\bf k}_4)d{\bf k}_1d{\bf k}_2d{\bf k}_3d{\bf
k}_4 \nonumber \\
\label{hamxishallow2b} +\frac{\sigma}{8(2\pi)^2}\int \Big \{
-({\bf k}_1\cdot{\bf k}_2)({\bf k}_3\cdot{\bf
k}_4)+\frac{h^2}{4}|{\bf k}_1+{\bf k}_2|^6+h^2k_3^4|{\bf
k}_1+{\bf k}_2|^2\nonumber \\
-k_1^4 \big [ (-1+3^{-1/2}) +(\frac{1}{2}-3^{3/2}2)h^2(|{\bf
k}_2+{\bf k}_3|^2+|{\bf k}_2+{\bf k}_4|^2+|{\bf k}_3+{\bf
k}_4|^2)\big ]
\nonumber \\
\times \xi_{{\bf k}_1}\xi_{{\bf k}_2}\xi_{{\bf k}_3}\xi_{{\bf
k}_4}
 \delta({\bf k}_1+{\bf k}_2+{\bf k}_3+{\bf k}_4)d{\bf
k}_1d{\bf k}_2d{\bf k}_3d{\bf k}_4.
\end{eqnarray}

\subsubsection{Nonzero gravity, $g\neq 0,$ and zero capillarity  $\sigma=0$}

\begin{eqnarray}
\label{hamxishallow0c} H_0=\frac{1}{2}\int \Big \{k^2 h |R_{\bf
k}|^2+g |\xi_{\bf k}|^2 \Big
\}d{\bf k}, \\
\label{hamxishallow1c} H_1=\frac{1}{4\pi}\int \Big[-({\bf
k}_1\cdot{\bf k}_2)R_{{\bf k}_1}R_{{\bf
k}_2}-\frac{gh}{6}(k_1^2+k_2^2+k_3^2)\xi_{{\bf k}_1}\xi_{{\bf
k}_2}\Big ] \nonumber\\
\times\xi_{{\bf k}_3}\delta({\bf k}_1+{\bf
k}_2+{\bf k}_3)d{\bf k}_1d{\bf k}_2d{\bf k}_3, \\
H_2=\frac{h}{8(2\pi)^2}\int \Big \{ 2({\bf k}_1\cdot{\bf
k}_2)^2-(k_1^2k_2^2+k_2^2k_1^2)\nonumber \\
+h^2\frac{1}{2}k_1^2k_2^2\big [|{\bf k}_1+{\bf k}_3|^2+|{\bf
k}_2+{\bf k}_3|^2+|{\bf k}_1+{\bf k}_4|^2+|{\bf k}_2+{\bf
k}_4|^2\big ]
\nonumber\\
+h^2 \frac{1}{4}k_1^2k_2^2\big [|{\bf k}_1+{\bf k}_2|^2+|{\bf
k}_3+{\bf k}_4|^2\big ]\Big \} \nonumber
\\
 \times R_{{\bf k}_1}R_{{\bf
k}_2}\xi_{{\bf k}_3}\xi_{{\bf k}_4}\delta({\bf k}_1+{\bf
k}_2+{\bf k}_3+{\bf k}_4)d{\bf k}_1d{\bf k}_2d{\bf k}_3d{\bf
k}_4 \nonumber \\
\label{hamxishallow2c} +\frac{g}{8(2\pi)^2}\int \Big \{
\frac{h^2}{4}|{\bf k}_1+{\bf k}_2|^4+h^2k_3^2|{\bf
k}_1+{\bf k}_2|^2\nonumber \\
-\frac{k_1^2h^2}{6} (|{\bf k}_2+{\bf k}_3|^2+|{\bf k}_2+{\bf
k}_4|^2+|{\bf k}_3+{\bf k}_4|^2)
\nonumber \\
\times \xi_{{\bf k}_1}\xi_{{\bf k}_2}\xi_{{\bf k}_3}\xi_{{\bf
k}_4}
 \delta({\bf k}_1+{\bf k}_2+{\bf k}_3+{\bf k}_4)d{\bf
k}_1d{\bf k}_2d{\bf k}_3d{\bf k}_4.
\end{eqnarray}

\section{Conclusion}

In conclusion, we found the optimal canonical variables for which
the water wave problem is well-posed in the approximation  which
keeps terms up to fourth order in the Hamiltonian. The important
question remain open if it is possible to make water wave
equations well-posed by proper choice of canonical transform for
higher-order corrections (fifth and higher order). We conjecture
that such optimal canonical variables exist in all orders of
nonlinearity.

%The authors thanks ... for helpful
%discussions.

Support was provided by   US Army Corps of Engineers, RDT\&E
Programm, grant DACA
 \# 42-00-C0044 (V.Z.),
  and to
 NSF grant \# NDMS0072803 (V.Z.).

%E-mail address: lushnikov@cnls.lanl.gov.

%\newpage

%\begin{references}

\newpage

\begin{figure}%[htbp]
\begin{center}
\includegraphics[width = 5 in]{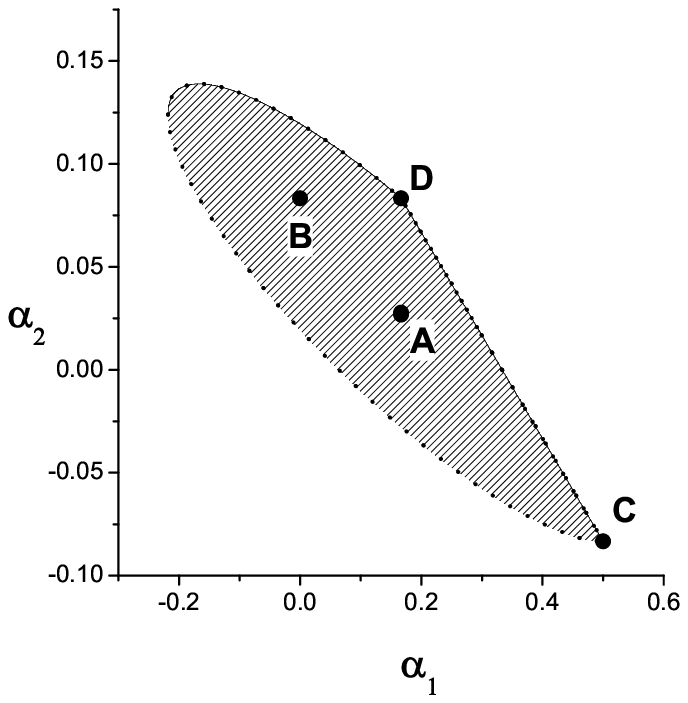}
%\vspace{0.5cm}
\caption{ Eqs.
$(\ref{hameqrxi1}),(\ref{hamxi0a})-(\ref{hamxi2a})$ with $g\neq
0$ are stable for any $(\alpha_1,\, \alpha_2)$ inside the filled
area. Point $A$ corresponds to the centre of ellipse,
$(\alpha_1,\, \alpha_2)=(\frac{1}{6}, \, \frac{1}{36})$. Point
$B$ corresponds to the most stable system
$(\ref{AB3dimensionless})$, $(\alpha_1,\, \alpha_2)=(0, \,
\frac{1}{12})$. Points $C$, $(\alpha_1,\, \alpha_2)=(\frac{1}{2},
\, -\frac{1}{12})$, and $D$, $(\alpha_1,\,
\alpha_2)=(\frac{1}{6}, \, \frac{1}{12})$, correspond to
intersections of the ellipse defined in
$(\ref{beta2cond})$ with the line $3\alpha_1+6\alpha_2=1$. lushnikovzakharovfig1.eps} %\label{fig:fig1}
\end{center}
\end{figure}
%

%\newpage

%
\begin{figure}%[htbp]
\begin{center}
\includegraphics[width = 5 in]{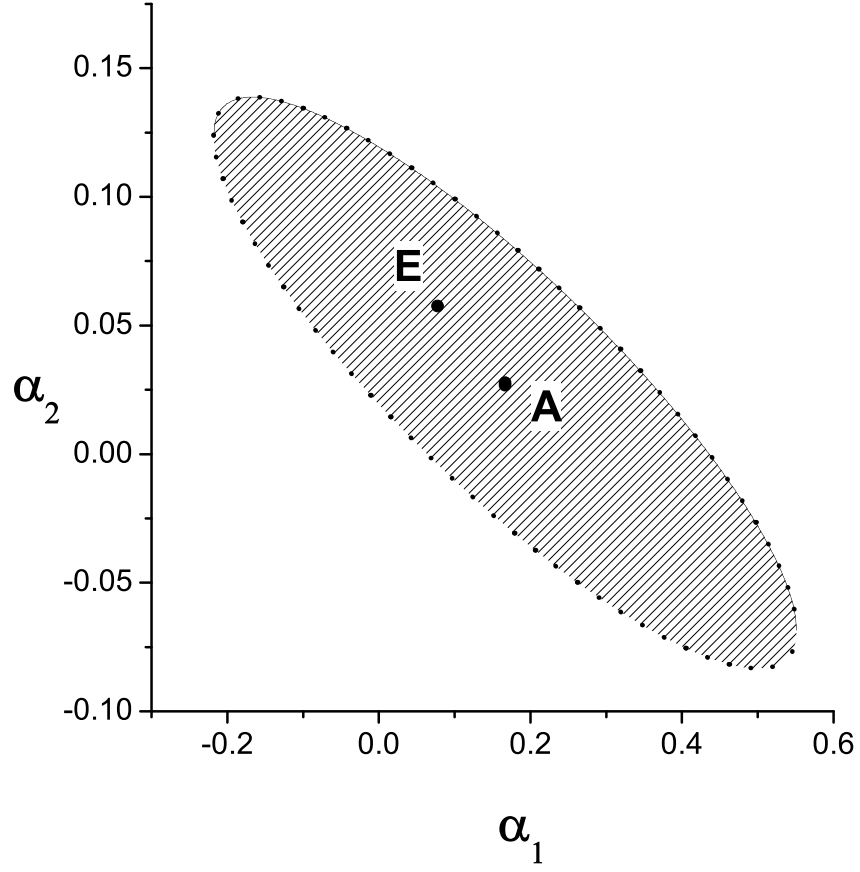}
%\vspace{0.5cm}
\caption{ Eqs.
$(\ref{hameqrxi1}),(\ref{hamxi0a})-(\ref{hamxi2a})$ with $g=0$
are stable for any $(\alpha_1,\, \alpha_2)$ inside the filled
area. The filled area is bounded by the ellipse defined in
$(\ref{beta2cond})$. Point $A$ corresponds to the centre of
ellipse, $(\alpha_1,\, \alpha_2)=(\frac{1}{6}, \, \frac{1}{36})$.
Point $E$, $(\alpha_1,\, \alpha_2)=(-\frac{1}{2}+3^{-1/2}, \,
\frac{1}{4}-3^{3/2})$, corresponds to the most stable system
$(\ref{AB3dimensionlessg0})$.
lushnikovzakharovfig2.eps} %\label{fig:fig1}
\end{center}
\end{figure}
%

%\newpage

%
\begin{figure}%[htbp]
\begin{center}
\includegraphics[width = 5 in]{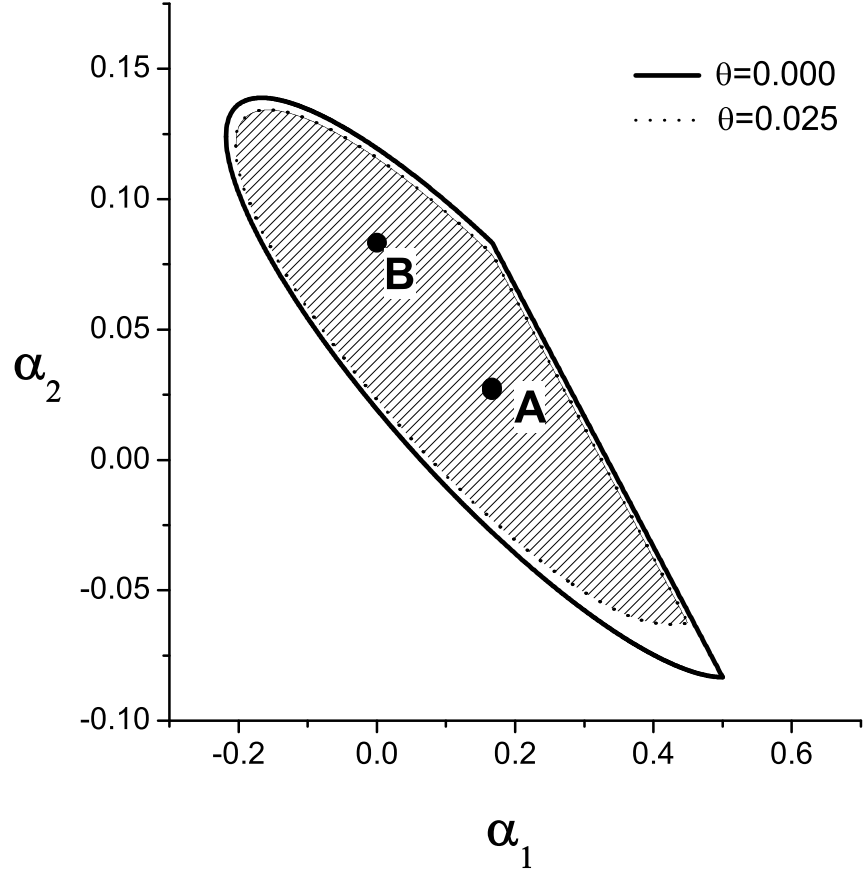}
%\vspace{0.5cm}
\caption{ Shrinking of  the stable region for small but nonzero
$\Theta$ for $g\neq 0.$ Solid curve corresponds to $\Theta=0$.
Area inside dotted curve corresponds to the stable region for
$\Theta=0.025.$ Dotted curve is obtained numerically for Eq.
$(\ref{AB3dimensionless})$ where we set as example
$O(\lambda\Theta)=-\lambda\Theta, \
O(\lambda^{-1}\rho\Theta)=\lambda^{-1}\rho\Theta, \
O(\lambda\rho\Theta)=-\lambda\rho\Theta.$ Points $A, \, B$ are
defined in Figure 1.
 lushnikovzakharovfig3.eps}
 %\label{fig:fig1}
\end{center}
\end{figure}
%

%\newpage

%
\begin{figure}%[htbp]
\begin{center}
\includegraphics[width = 5 in]{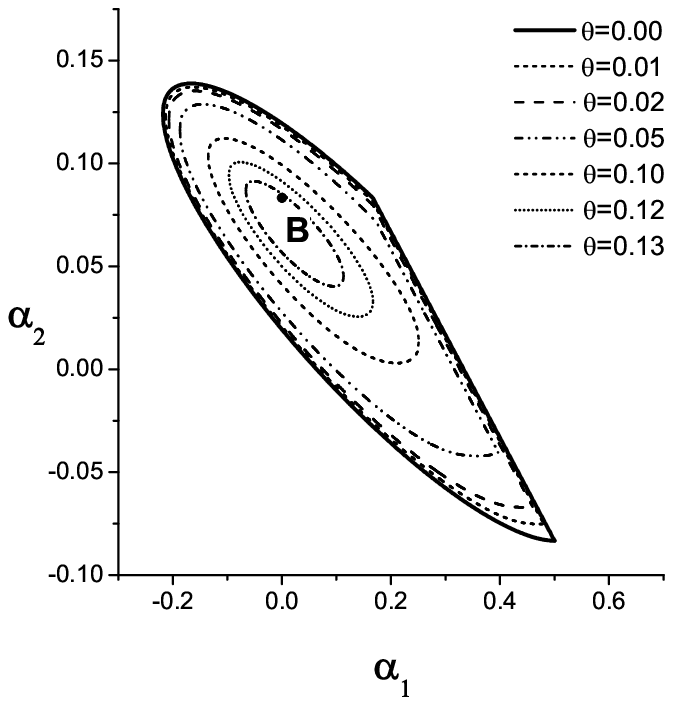}
%\vspace{0.5cm}
\caption{ A series of curves corresponding to shrinking of the
stable region as a function of $\Theta$ for $g\neq 0$. For
$\Theta<0.1$, the width of region between solid curve ($\Theta=0$)
and curves with $\Theta\neq 0$ scales approximately as $\Theta$.
For $\Theta>0.1$, the stable region  quickly shrinks to zero as
$\Theta$ approaches $\simeq 0.1389$.  Similar to Figure 3, all
curves are calculated numerically with assumptions
$O(\lambda\Theta)=-\lambda\Theta, \
O(\lambda^{-1}\rho\Theta)=\lambda^{-1}\rho\Theta, \
O(\lambda\rho\Theta)=-\lambda\rho\Theta.$
lushnikovzakharovfig4.eps} %\label{fig:fig1}
\end{center}
\end{figure}
%

%\newpage

%
\begin{figure}%[htbp]
\begin{center}
\includegraphics[width = 5 in]{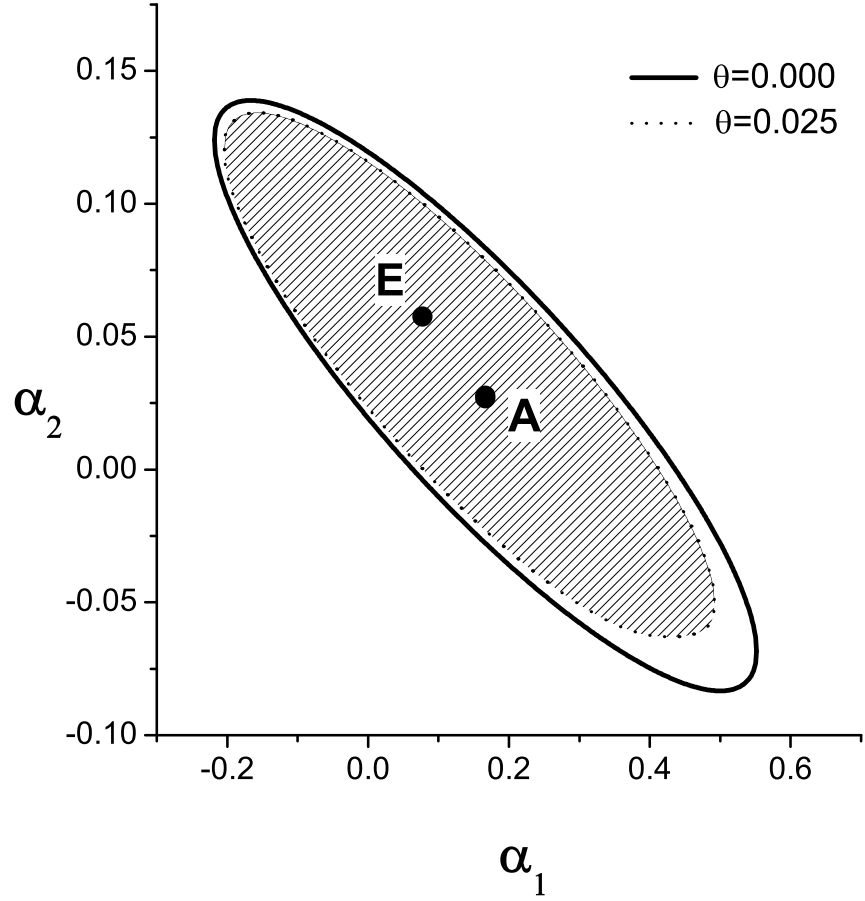}
%\vspace{0.5cm}
\caption{ Shrinking of  the stable region for small but nonzero
$\Theta$ for $g=0.$ Solid curve corresponds to $\Theta=0$. Area
inside dotted curve corresponds to the stable region for
$\Theta=0.025.$ Dotted curve is obtained numerically for Eq.
$(\ref{AB3dimensionless})$ where we set as example
$O(\lambda\Theta)=-\lambda\Theta, \
O(\lambda^{-1}\Theta)=\lambda^{-1}\Theta.$ Points $A, \, E$ are
defined in Figure 2.
lushnikovzakharovfig5.eps} %\label{fig:fig1}
\end{center}
\end{figure}
%

%\newpage

%
\begin{figure}%[htbp]
\begin{center}
\includegraphics[width = 5 in]{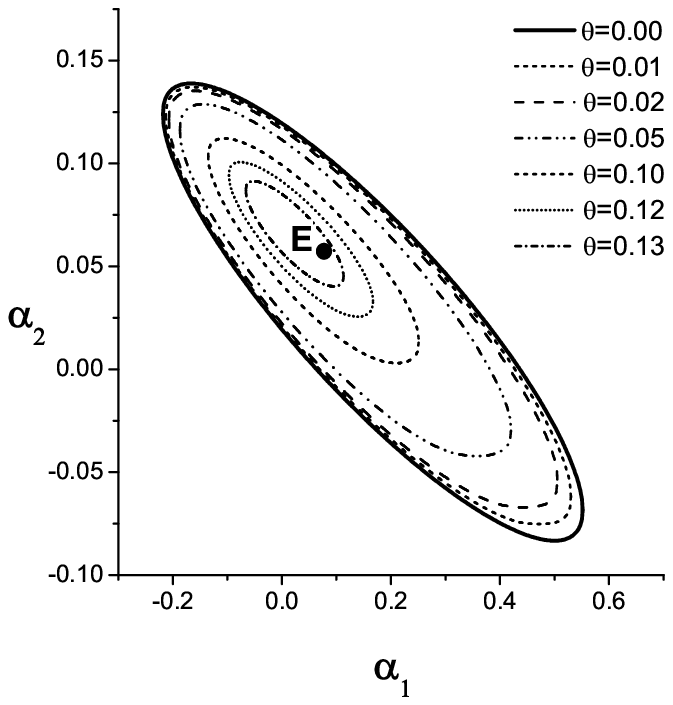}
%\vspace{0.5cm}
\caption{ A series of curves corresponding to shrinking of the
stable region as a function of $\Theta$ for $g=0$. For
$\Theta<0.1$, the width of region between solid curve ($\Theta=0$)
and curves with $\Theta\neq 0$ scales approximately as $\Theta$.
For $\Theta>0.1$, the stable region  quickly shrinks to zero as
$\Theta$ approaches $\simeq 0.1389$.  Similar to Figure 5, all
curves are calculated numerically with assumptions
$O(\lambda\Theta)=-\lambda\Theta, \
O(\lambda^{-1}\Theta)=\lambda^{-1}\Theta.$
lushnikovzakharovfig6.eps} %\label{fig:fig1}
\end{center}
\end{figure}

\end{document}